\documentclass[preprint2]{pp7}
\usepackage{times}
\usepackage{amsmath}

\usepackage[backref,breaklinks,pdfnewwindow=true,colorlinks=true,citecolor=blue]{hyperref}
\usepackage[all]{hypcap}

\usepackage{etoolbox}

\usepackage[usenames,dvipsnames]{xcolor}
\usepackage{color}

\begin{document}

\title{\textbf{\LARGE Kinematic Structures in Planet-Forming Disks}}

\author {\textbf{\large Christophe Pinte}}
\affil{\small\em School of Physics and Astronomy, Monash University, Clayton Vic 3800, Australia}
\affil{\small\em Univ. Grenoble Alpes, CNRS, IPAG, F-38000 Grenoble, France}
\author {\textbf{\large Richard Teague}}
\affil{\small\em Center for Astrophysics \textbar{} Harvard \& Smithsonian, 60 Garden Street, Cambridge, MA 02138, USA}
\author {\textbf{\large Kevin Flaherty}}
\affil{\small\em Department of Astronomy and Department of Physics, Williams College, Williamstown, MA, USA}
\author {\textbf{\large Cassandra Hall}}
\affil{\small\em Department of Physics and Astronomy, The University of Georgia, Athens, GA 30602, USA}
\affil{\small\em Center for Simulational Physics, The University of Georgia, Athens, GA 30602, USA}
\author {\textbf{\large Stefano Facchini}}
\affil{\small\em European Southern Observatory, Karl-Schwarzschild-Str. 2, 85748 Garching bei München, Germany}
\affil{\small\em Dipartimento di Fisica, Universitá degli Studi di Milano, via Celoria 16, 20133 Milano, Italy}
\author {\textbf{\large Simon Casassus}}
\affil{\small\em Departamento de Astronom\'ia, Universidad de Chile, Casilla 36-D, Santiago, Chile}

\begin{abstract}
\baselineskip = 11pt
\leftskip = 0.65in
\rightskip = 0.65in
\parindent=1pc
{\small The past 5 years have dramatically changed our view of the disks of gas and dust around young stars. Observations with the Atacama Large Millimeter/submillimeter Array (ALMA) and extreme adaptive optics systems have revealed that disks are dynamical systems. Most disks contain resolved structures, both in gas and dust, including rings, gaps, spirals, azimuthal dust concentrations, shadows cast by misaligned inner disks, as well as deviations from Keplerian rotation. The origin of these structures and how they relate to the planet formation process remain poorly understood.
Spatially resolved kinematic studies offer a new and necessary window to understand and quantify the physical processes (turbulence, winds, radial and meridional flows, stellar multiplicity, instabilities) at play during planet formation and disk evolution.
Recent progress, driven mainly by resolved ALMA observations, includes the detection and mass determination of embedded planets, the mapping of the gas flow around the accreting planets, the confirmation of tidal interactions and warped disk geometries, and stringent limits on the turbulent velocities.
In this chapter, we will review our current understanding of these dynamical processes and highlight how kinematic mapping provides new ways to observe planet formation in action.
 \\~\\~\\~}
\end{abstract}


\section{\textbf{Introduction}}

Since the discovery of the first exoplanet, 51 Pegasi b \citep{Mayor1995}, the number of confirmed exoplanets has steadily increased, jumping from around 1000 at the time of Protostars and Planets VI to almost 5000 today. These detections have revealed a stunning diversity in the properties of exoplanets, with planet masses and orbital separations spanning more than five orders of magnitudes \citep{Winn2015}, a large variety in planet density, and hence composition \citep{Kaltenegger2017}, as well as in planet atmosphere composition \citep{Madhusudha2019}.
They demonstrated that the planet formation process is ubiquitous, efficient and robust. But they also challenged our understanding of the pathways of planet formation. Part of this diversity must originate from the  initial few million years of these planets lifetime, during the building phase of the disk from the molecular cloud, its evolution and interaction with surrounding environment, and finally its dissipation. The relative role of these phases on the final architecture of the planetary systems is unknown.
The detection and characterisation of young planets is critical for constraining the mechanisms and timescales of planet formation and migration, as well as the chemical composition of the material accreted onto the planets.
Young ($\lesssim$ 10\,Myr) planets have remained elusive however, with only about 10 that have been detected to date (NASA exoplanet archive).

The most prolific methods to detect mature planets (transits and radial velocities, see chapter by Lissauer et al.) are limited by the magnetic activity of young stars which display photometric and spectroscopic variability on a multitude of timescales, with amplitudes that can be orders of magnitude larger than those induced even by massive planets. Only 3 planets have been found around young stars from radial velocity searches: CI~Tau~b \citep{Johns-Krull2016, Flagg2019}, V830~Tau~b \citep{Donati2015, Donati2016, Donati2017} and TAP 26 b \citep{Yu2017}, but these detections have been challenged \citep{Damasso2020, Donati2020}. For close to edge-on configurations required for transit observations, detections are near impossible due the presence of the optically thick disk blocking the view. The only transiting planet detected to date around a star younger than 10\,Myr \citep[K2-33~b,][]{David2016} is surrounded by an optically thin debris disk.

Near infrared direct imaging surveys have revealed planet-mass companions at wide separations (few to tens of Jupiter masses, tens to hundreds of astronomical units from host stars). For stars younger than 10~Myr, most of these companions have been found around brown dwarfs, where the star/planet contrast is more favorable \citep{Chauvin2004, Luhman2006, Lafreniere2008, Bejar2008, Todorov2010, Fontanive2020}.
The characterisation of these wide separation candidates remains difficult, as extinction (for instance by a highly inclined disk) can strongly dim the object \citep{Wu2017, Christiaens2021}.

The new generation of extreme adaptive optics (AO) instruments (MagAO, SCEXAO, GPI, VLT/SPHERE), despite their high contrast and angular resolution, have demonstrated the difficulty in imaging planets still embedded in their native disk (see chapters by Benisty et al. and Currie et al.). In the near-infrared, the disk is optically thick, even at hundreds of astronomical units from the star, and extincts any potential planet \citep[e.g.,][]{Sanchis2020}. The disk also scatters stellar light, and while it is hundreds of times fainter than the star itself, it will outshine any embedded planet.
Detections have been reported in several disks: HD~100546 \citep{Quanz2013a, Brittain2014, Quanz2015, Currie2015}, LkCa~15 \citep{Kraus2012, Sallum2015}, HD~169142 \citep{Quanz2013b, Biller2014, Reggiani2014}, and MWC 758 \citep{Reggiani2018}. Yet, most of the detections to date have been subsequently challenged \citep[e.g.,][]{Thalmann2015, Thalmann2016, Rameau2017, Ligi2018, Currie2019}. The remarkable exception is PDS~70 where a system of two planets was imaged in the cavity of the transition disk \citep{Keppler2018, Muller2018, Wagner2018, Christiaens2019, Haffert2019, Wang2020, Wang2021, Zhou2021}, very likely because the central cavity is strongly depleted which alleviates the previously mentioned issue of optical depth and disk/planet brightness contrast.

While the direct detection of young exoplanets remains rare, even after years of survey with dedicated instruments, near-IR AO systems have revealed a wealth of stunning images of disks. Together with ALMA, they have pushed the boundaries of our understanding and have changed the field of planet formation from mostly theoretical to observation driven. They revealed that most if not all disks are highly structured \citep{Andrews2020}, with rings/gaps \citep[e.g.][]{ALMA_HLTau, Andrews2016, Muro-Arena2018, Long2018, Andrews2018, Huang2018b, vanTerwisga2018, Avenhaus2018, Villenave2019, Perez2020c}, central cavities \citep[e.g.][]{Casassus2013, vanderPlas2017, vanderPlas21017b, Pinilla2017, Pinilla2018, Pinilla2019, Long2018, vanderMarel2018, Kudo2018, Facchini2020}, spirals \citep[e.g.][]{Hashimoto2011, Garufi2013, Perez2016, Benisty2015, Benisty2017, Stolker2017, Huang2018c}, azimuthal asymmetries/arcs \citep[e.g.][]{van-der-Marel2013, Casassus2013, Perez2014, Kraus2017, Dong2018, Perez2018b, Cazzoletti2018, Isella2018}. Multiple types of structures can co-exist in the same system, especially when comparing different tracers. The most spectacular examples are disks showing spirals in scattered light while exhibiting rings at sub-millimeter wavelengths, such as MWC~758 \citep{Benisty2015,Dong2018} and HD~135344~B \citep{Stolker2017,Cazzoletti2018}. While this could trace multiple processes operating simultaneously in the disk, this is likely indicative of the varying response to same physical mechanism as a function of the dust grains sizes and altitude.

The most exciting explanation to these observations is that disks already host unseen young, nearly formed, planets \citep[e.g.][]{Zhu2011, Pinilla2012, Gonzalez2012, Dipierro2015, Rosotti2016, Bae2017}. Alternative mechanisms have been suggested and have been shown to produce comparable structures (see chapter by Bae et al. for a detailed discussion). For instance, rings and gaps in the dust distribution -- the most commonly detected substructures -- can be the result of condensation fronts \citep{Kretke2007, Saito2011, Ros2013, Zhang2015}, dust sintering \citep{Okuzumi2012},
various (magneto-) hydrodynamical instabilities, such as the magneto-rotational instability \citep[MRI;][]{Simon2014, Flock2015, Bethune2016, Riols2019, Riols2020}, zonal flows \citep{Uribe2011}, thermal wave instabilities \citep{Ueda2021b}, dust instability \citep{Loren-Aguilar2015} and self-induced dust traps \citep{Gonzalez2017}, baroclinic instabilities \citep{Klahr2003}, secular gravitational instability \citep{Takahashi2016}, counter-rotating infall \citep{Vorobyov2016} or radially variable magnetic disk winds \citep{Suzuki2016, Suriano2017, Suriano2018, Suriano2019}.
Additional signatures of planet-disk interactions are critical to confirm the presence of a planet.

\begin{figure*}[!h]
  \includegraphics[width=\textwidth]{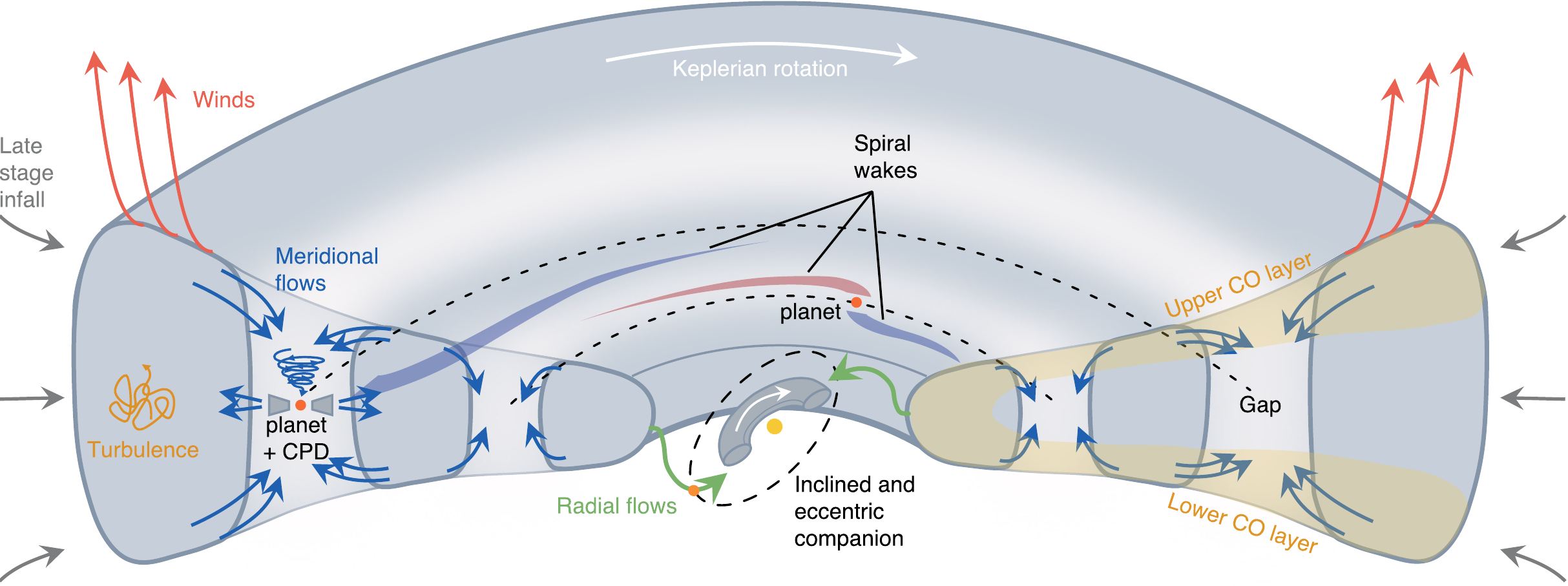}
  \caption{Schematic view of the expected kinematic signatures in a planet forming disk. Planets generate Lindblad resonances and buoyancy spirals,  and can carve gaps that will be associated with meridional flows. At small scales a circumplanetary disk can form around the planet.  Inner companions on inclined and/or eccentric orbits can tilt an inner disk and generate near free-fall radial flows while clearing a central cavity. Unstructured turbulent motions can locally broaden the line width.  Disks can interact with their environment via winds that likely remove the disk angular momentum, and late stage infall or flyby tidal interactions that may impact the mass, angular momentum and chemical history of the disk. High spectral resolution molecular line observations, for instance probing the warm CO layer, offer a way to map the gas flows associated with these processes.}
\end{figure*}

Molecular line observations of the gaseous component offer a complementary window to distinguish between these competing theories. Several disks show substructures in molecular emission suggestive of surface density perturbations \citep[e.g.][]{Isella2016, Teague2017, Huang2018a, Seongjoong2020}. Distinguishing these features from chemical effects \citep{Oberg2015, Bergin2016, Cazzoletti2018b, Kastner2018, MAPS1, MAPS3} remains difficult however, and care must be taken to disentangle density structures from temperature effects, for instance due to shadowing, or artefacts due to continuum subtraction \citep[e.g.][]{Rab2020, Rosotti2021}.

One of the revolutions of ALMA is its ability to map molecular emission from disks at high spatial and spectral resolutions. This offers an exciting possibility to map the kinematic structures, bypassing the complexities of having to account for the molecular excitation and underlying chemical structures. Such studies have led to key results that are progressively transforming our understanding of disks and planet formation, including:
\begin{itemize}
    \item accurate determination of central stellar masses,
    \item the mapping of tidal interactions between multiple stars and flybys,
    \item the role of inner companion in carving the large cavities in some transition disks,
    \item constraints on the level of turbulent motions and the $\alpha$ viscosity, a fundamental parameter which determines how angular momentum transport occurs in a disk,
    \item the extraction of maps of the temperature structure of disks,
    \item the detection of molecular disk winds,
    \item indications of gravitational instabilities,
    \item and the kinematic detections of embedded planets.
\end{itemize}
This field remains in its infancy, and there are only a few disks that have been observed with high enough integration time and spectral resolution to conduct these studies. An important aspect of this chapter is the discussion of these new advances brought about by ALMA, which are continually developing.

We first describe the methods to map and reconstruct the velocity fields in disks and their surrounding environment in \S 2. Next we discuss the progress in mapping the large scale flows in \S 3, and discuss the interplay between planet and star formation in \S 4 with emphasis on multiple systems, misalignments and warps. In \S 5, we summarise our current understanding of planet-disk interactions and discuss the kinematic signatures of embedded planets, and the implication of recent detections. We conclude in section \S 6 by discussing the perspectives that will be offered with future instruments.

\section{Kinematic observations of disks: methods}

In most of this chapter, we will focus on line observations in the sub-millimeter regime, in particular observations with the ALMA interferometer. The techniques are applicable to any line data that is both spatially and spectrally resolved, for instance with integral field units (IFUs). The end product of line observations is a data cube, with three axes (and eventually a polarisation axis): two spatial axes corresponding to the sky coordinate, and a frequency axis. The frequency can be converted to line-of-sight velocity via calculation of the Doppler shift of the line from its rest frequency. Data cubes are composed of channel maps, showing the spatial emission morphology at a specific frequency (or velocity).

The large collecting area of the ALMA interferometer enables the imaging of protoplanetary disks at high spectral resolution. The maximal spectral resolution of 30.5~kHz translates into a velocity resolution of $26~\mathrm{m\,s}^{-1}$ and $40~\mathrm{m\,s}^{-1}$ for the $^{12}$CO J=3-2 and J=2-1 lines for instance. This probes a new range of physical processes that were inaccessible to the previous generation of (sub-)millimeter instruments. The high spectral resolution required to image kinematic structures also means that the available bandwidth is much narrower than for continuum observations, and line observations require long integration times, even with ALMA, and preclude the use of the longest baselines.  As such, much of the substructure observed in disks has been detected in the continuum so far. But reasonable compromises on spatial resolution still allow for a fine mapping of the disk velocity fields. For instance, in band 7 at 330\,GHz (e.g.~$^{12}$CO J=3-2), a brightness sensitivity of 5\,K and velocity resolution of 26\,m\,s$^{-1}$ can be reached in $\approx$ 30\,min and 8\,h on source time at a spatial resolution of 0.2'' and 0.1'', respectively. While the number of sources where kinematics studies can be performed remains limited, it seems that (not surprisingly), most if not all of the sources display kinematic substructures on top of a smooth Keplerian structure when data sets with sufficient spectral resolution and integration time are available.

\subsection{Velocity fields in protoplanetary disks}
\label{sec:velocity_fields}

\begin{figure*}
  \includegraphics[width=\textwidth]{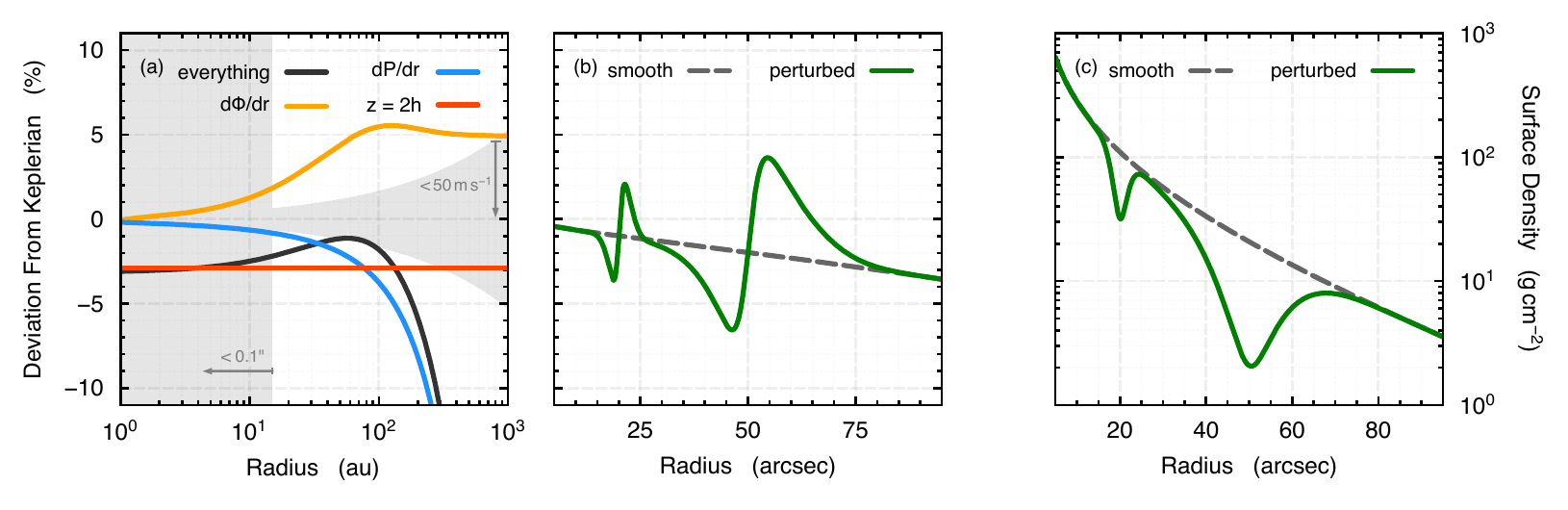}
  \caption{A comparison of different rotation curves. Panel (a) compares the fractional difference relative to Keplerian rotation when a non-zero emission height ($z = h = 0.1r$, red), a smooth radial negative pressure gradient with an exponential taper (blue) and disk self-gravity (where M$_\mathrm{disk} =0.1~$M$_*$, and $\Sigma(r) \propto r^{-1}$, orange) are included. The black line shows the resulting deviation when all components are considered together. The shaded area represents the current limitations in terms of spatial (0.1'', assuming 150\,pc ) and spectral resolution (50\,m\,s$^{-1}$). Panel (b) shows the impact of a perturbed pressure profile due to gaps in the gas surface density shown in panel (c), without considering the contribution of the self-gravity term. \label{fig:azimuthal_velocities}
  }
\end{figure*}

It is insightful to compare the velocity resolution ALMA can reach with the expected velocities from key physical processes happening in disks.
Protoplanetary disks are mostly rotating at Keplerian velocities and in vertical hydrostatic equilibrium, meaning that there is no bulk vertical or radial motions. Assuming the gravity is dominated by the central star and neglecting pressure gradients, the midplane Keplerian velocity is given by the balance between the gravity and centrifugal forces:
\begin{equation}
v_{\phi,\,\mathrm{K}}(r) = \sqrt{\frac{G M_*}{r}} \approx 3.0 \sqrt{ \frac{M_*}{M_\odot}} \sqrt{\frac{100\mathrm{au}}{r} } \mathrm{km\,s}^{-1}
\label{eq:orbital_velocity_keplerian}
\end{equation}

The vertical extent of the disk reduces the gravitational force from the star as well as its radial projection, leading to an orbital velocity
\begin{equation}
  \frac{v_\phi(r,z)^2}{r} = \frac{G M_*}{r^2 + z^2}\ \times \frac{r}{\sqrt{r^2 + z^2}}
  \label{eq:velocity_vertical_dependence}
\end{equation}
For a vertically isothermal disk in hydrostatic equilibrium with a typical scale height ratio $h/r$ of 0.05, this corresponds to a difference of velocity of $\approx 5\%$ between the disk midplane and 4 scale heights, the typical altitude of $^{12}$CO emission.
At a distance of 100\,au from the star, such difference amounts to $\approx 170$\,m\,s$^{-1}$.

In addition to the gravity from the star, any extra force exerted on the gas will affect the orbital velocities.
Including the pressure gradient, the previous equation becomes
\begin{equation}
  \frac{v_\phi(r,z)^2}{r} = \frac{G M_* r}{(r^2 + z^2)^{3/2}} + \frac{1}{\rho_\mathrm{gas}}\frac{\partial P_\mathrm{gas}}{\partial r}
  \label{eq:pressure_gradient}
\end{equation}

The pressure gradient term is in general complex and depends on the details of the disk density and thermal structures \citep[e.g.][]{Rab2020}. But we can get an estimate of its impact on the orbital velocities under simple assumptions on the disk structure. If the pressure is a monotonically decreasing function, $\frac{\partial P_\mathrm{gas}}{\partial r}$ is negative, and the gas orbits at sub-Keplerian velocities.
Assuming power-laws for the surface density $\Sigma (r) = \Sigma_0 (r/r_0)^{-p}$ and a vertically isothermal temperature $T = T_0 (r/r_0)^{-q}$ (and hence a power $P = P_0 (r/r_0)^{-s}$ for the midplane pressure, with $s = p +q/2 + 3/2$), the midplane azimuthal velocities become
\begin{equation}
  v_\phi(r) = v_{\phi,\,\mathrm{K}}(r) \sqrt{1-s \left(\frac{h}{r}\right)^2}
\end{equation}
where $h$ the local scale height. With a surface density index $p=1$, and a midplane temperature profile index $q = 0.5$, the typical deviations from Keplerian velocity are of the order of 1\,\%. While this term is small, it is responsible for the gas drag on dust grains. As dust grains are not pressure supported, they orbit at Keplerian velocities and feel a headwind from the gas. They lose angular momentum and drift inwards. For a smooth disk, such a velocity deviation is in practice not detectable, in particular because the true Keplerian velocity (and central mass star) is only known to a limited accuracy. If the pressure gradient becomes steeper (for instance at the outer disk where the surface density tapers off, or at the edges of a gap in the gas surface density), this pressure term becomes significant and relative measurements are possible.

Finally, considering the disk self-gravity, the orbital velocities can be expressed as
\begin{equation}
\frac{v_\phi(r,z)^2}{r} = \frac{G M_* r}{(r^2 + z^2)^{3/2}} + \frac{1}{\rho_\mathrm{gas}}\frac{\partial P_\mathrm{gas}}{\partial r} + \frac{\partial \phi_\mathrm{gas}}{\partial r}
\label{eq:orbital_velocity_total}
\end{equation}
where $ \phi_\mathrm{gas}$ is the gravitational potential of the disk. The  $\frac{\partial \phi_\mathrm{gas}}{\partial r}$ term can be computed numerically and depends on the details of the disk surface density \citep[e.g.][]{Bertin1999,Veronesi2021}. In the case of an infinitely thin disk extending to infinity with a surface density $\Sigma (r) \propto r^{-1}$, a simple analytical expression can be derived \citep{Mestel1963,Lodato2007}:
\begin{equation}
\frac{\partial \phi_\mathrm{gas}}{\partial r} = 2\pi G \Sigma (r)
\label{eq:grav_potential}
\end{equation}
where $G$ is the gravitational constant.
This term is positive, leading to super-Keplerian rotation, and because it decreases slower with radius than the stellar gravity,
the relative difference between the orbital and Keplerian velocities will progressively increase with distance. Assuming that the disk mass remains small compared to the central mass, the velocities in the outer disk will differ from the velocities extrapolated from the inner part by a ratio $\approx 1/2  \times M_\mathrm{disk} / M_*$. This means that a massive disk with $M_\mathrm{disk} =0.1 M_*$ could speed up the gas flow by a few percents. This is potentially measurable if the velocities can be accurately measured over a wide range of radii in the disk. Figure~\ref{fig:azimuthal_velocities} illustrates the impact of these various terms on the gas orbital velocities.

On top on the azimuthal flows, disks also experience a slow viscous evolution. Assuming that the viscosity can be described by a viscosity parameter $\alpha$ \citep{Shakura73}, the accretion velocities can be written as
\begin{equation}
  |v_\mathrm{r, \mathrm{acc}}| \sim  \alpha \left(\frac{h}{r}\right)^2 v_\mathrm{Kep} \lesssim 10^{-4}\, v_\mathrm{Kep}
\end{equation}
This term is in general too small to be measured. Under the assumption that accretion is constant $\dot{M} = 2\pi r \Sigma v_{r, \mathrm{acc}}$, and $v_{r, \mathrm{acc}} \propto 1/\Sigma$, this diffusive velocity can be significantly increased to measurable levels if the surface density drops, for instance inside a gap or central cavity \citep[e.g.][]{Rosenfeld2014}.

The molecular emission is set by the global velocity, but also by the local linewidth. A key quantity describing the gas Brownian motions is the sound speed given by
\begin{equation}
  c_\mathrm{s} = \sqrt{\frac{k_\mathrm{B}T_\mathrm{gas}}{\mu m_\mathrm{h}}} \approx 300\, \sqrt{\frac{T_\mathrm{gas}}{25\mathrm{K}}}\, \mathrm{m\,s}^{-1}
\end{equation}
where $k_\mathrm{B}$ is the Boltzmann constant, $\mu=2.3$ is the mean molecular weight (for a fully molecular gas of interstellar composition) in units of the proton mass $m_\mathrm{h}$ and $T_{\rm gas}$ the gas temperature.
For rotational transitions of molecules in the sub-millimeter regime, line broadening is dominated by the thermal and turbulent motions. The corresponding local line profiles are Gaussian, with the line width set by the root mean square (RMS) of the Maxwellian velocity distributions:
\begin{equation}
  \Delta v = \sqrt{\frac{2k_\mathrm{B} T_\mathrm{gas}}{m_\mathrm{mol}} + \delta v_\mathrm{turb}^2}
\end{equation}
where $m_\mathrm{mol}$ is the mass of the considered molecule. In particular, in the absence of turbulence, the thermal RMS velocities for a molecule like CO is $\sqrt{ m_\mathrm{CO} / \mu m_\mathrm{H}} \approx 3.4$ times smaller than the local sound speed, \emph{i.e.} of the order of 120\,m\,s$^{-1}$ for a gas temperature of 25\,K.

The sound speed is also the typical velocity that is expected for the base of a molecular wind in the outer parts of the disk, where the gas is weakly gravitationally bound.

Embedded bodies and external companions or flybys can generate tidal forces which can disturb the Keplerian flow. No simple estimate of the velocity deviations exist in the general case as the tidal interactions depend on the mass ratio, orbital parameters, and disk physical structure (in particular temperature and viscosity). For an embedded planet on a circular orbit with a mass lower than the thermal mass  $m_\mathrm{th} = (h_p/r_p)^3 M_*$, where $h_p$ is the local hydrostatic scale height at the orbital radius of the planet $r_p$, \cite{Goodman2001} developed a linear theory and showed that the velocity perturbations scale as $m_p/m_\mathrm{th}$, where $m_p$ is the planet mass. Semi-analytical calculations \citep{Bollati2021} showed that a planet of 0.5\,M$_\mathrm{Jup}$ planet at 100\,au around a 1\,M$_\odot$ star will generate deviations from Keplerian rotation of the order of 5\% for $h/r = 0.1$, corresponding to $\approx$ 150\,m\,s$^{-1}$.

\subsection{The stratification of the molecular emitting layer }

The most abundant molecule in disks (H$_2$) does not have a permanent dipole moment, and is transparent in the vast majority of the disk. To probe the gas structure, measurements rely on the emission of sub-millimeter spectral lines in the rotational transitions of less abundant molecules.

Protoplanetary disks host steep radial and vertical gradients in density, radiation field and temperature. A remarkable consequence of this structure is the vertical chemical stratification of disks \citep[e.g.][]{Henning2013,Dutrey2014_PPVI}, with a cold midplane where most molecules are frozen on the dust grains, a hot surface directly illuminated by the central star, where UV photons dissociate molecules, and an intermediate warm molecular layer where most molecules live in abundance \citep{vanZadelhoff2001,Aikawa2002,Dartois2003,Semenov2008}. Emission from each molecule will predominantly trace regions where the abundance of that molecule is largest, and, as each molecule requires slightly different physical conditions to thrive, this enables us to trace distinct vertical regions within the disk. For abundant  molecules (e.g. $^{12}$CO), the molecular layer is optically thick and the actual emission layer probed by sub-millimeter observations is even narrower as the $\tau=1$ surface is reached quickly, while for less abundance molecules, the molecular emission will probe the full line-of-sight column. The altitude of the emitting layers depends on the disk physical, thermal and chemical structures.  With a careful choice of molecules and/or isotopologues, one can trace a range of altitudes within the disk. Typical values are 3-4 hydrostatic scale heights for $^{12}$CO and about one scale height for $^{13}$CO and C$^{18}$O \citep[e.g.][]{Rosenfeld2014,Pinte2018a,MAPS4}.
Interestingly, scattered light images in the near-infrared appear to probe a similar layer, between 2 and 3 scale heights, \citep{Ginski2016,Stolker2016,Avenhaus2018}, and comparison of CO maps with scattered light images allows to search for the gas counter part of the sub-structures seen in the small dust phase.

The formation of emission lines from protoplanetary disks is the combination of the local line excitation conditions and the global Keplerian shear.
The local line profile, and level populations  (and hence the emissivities and absorption) are set by the physical and chemical conditions at any point in the disk: molecular abundance, gas temperature, density of collision partners, turbulence level and local radiation field all set the spectral and angular distribution of the radiation emerging from each point in the disk. Note that this is in principle a non-local problem as the radiation field at any point is an integral  of the radiation field from any other point, which in turn depends on the statistical equilibrium. The gas densities at which the low-J rotational lines are formed (or where they reach an optical depth $\approx 1$) are typically higher than the critical densities, meaning that collisional coupling will dominate the level population and the line emission will be at the local thermodynamic equilibrium (LTE, e.g. \citealp{Pavlyuchenkov2007}).
The global emission detected by an observer is then the result of a formal integration of the radiative transfer equation, where the emissivities and absorptions are defined by the individual contributions from each point shifted by the appropriate Doppler shift to account for the orbital motion of the gas.

Local line width and Doppler shift have different effects on the observed emission (ordered flows shift the line center while small scale, turbulent like motions broaden the line), and it is possible to reconstruct both the local conditions and the velocity field with various methods we detail below. In practice, we can only observe the line-of-sight projection of the velocity fields, and the finite spectral and spatial resolution smear out some of the details. Partly optically thin lines additionally sample a potentially large range of altitudes and radii adding to the complexity of any analysis.

\subsection{Channel maps and moment maps}
\label{sec:channel_maps}

\begin{figure*}[!h]
  \includegraphics[width=\textwidth]{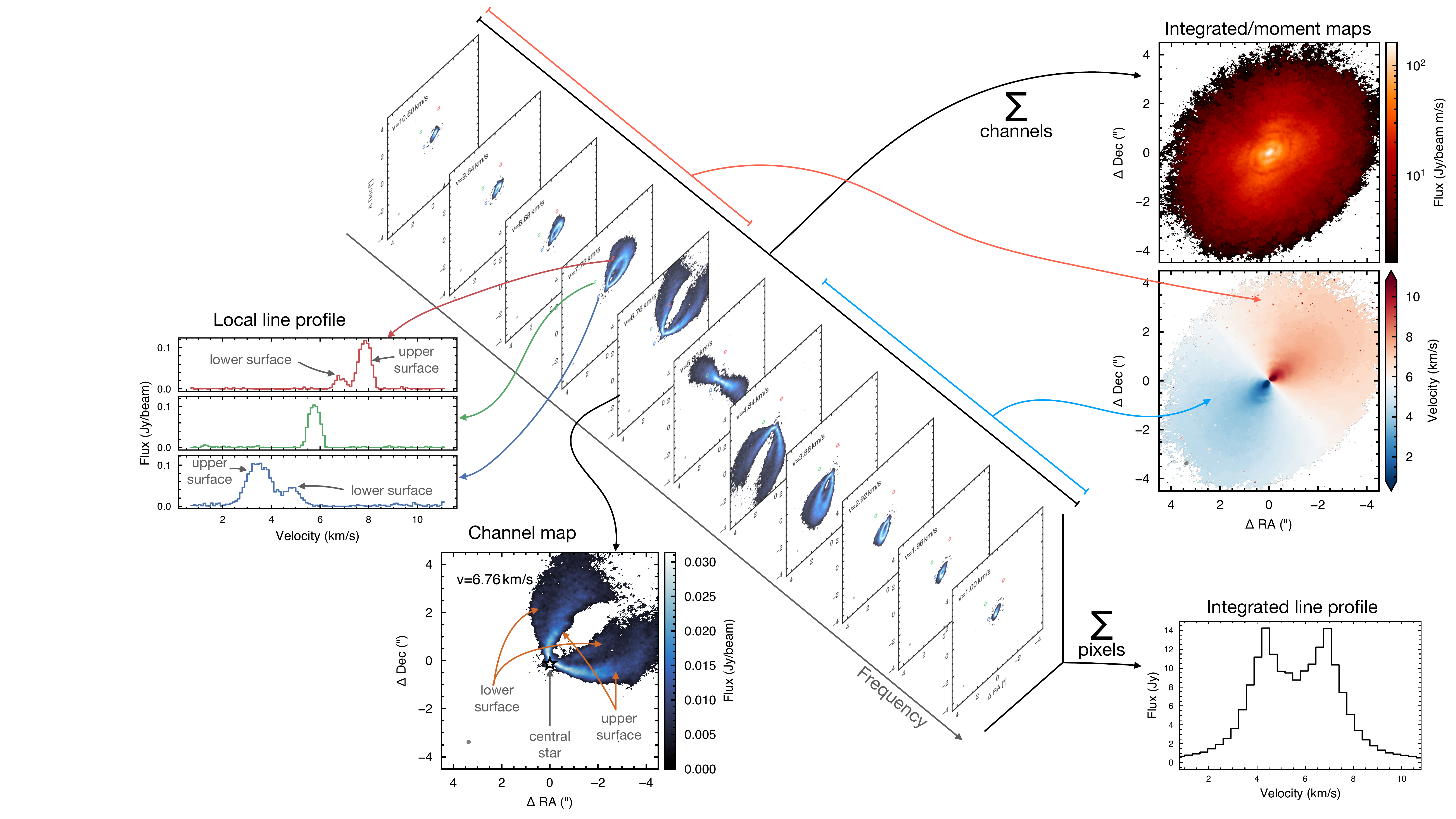}
  \caption{Multiple ways to view a molecular data cube, here using the $^{12}$CO J=2-1 ALMA observations of HD~163296 \citep{Isella2016,Andrews2018}. Clockwise from bottom: at a given velocity, a single map can be extracted; local line profiles can be computed by integrating spatially the flux over a selected region; moment maps are computed by integrating along the velocity axis or fitting line profiles for each pixel; a global  spectrum is computed by integrating spatially over the map. \label{fig:channel_maps}}
\end{figure*}

At a given frequency, line emission is concentrated along an isovelocity curve, \emph{i.e.} the region where the projected velocity of the emitting surface is constant (and corresponds to difference between the selected frequency and the rest frequency of the transition). Because of the Keplerian rotation of the disk, this isovelocity curve follows a typical ``butterfly'' pattern \citep{Horne1986,Beckwith1993}.
While pre-ALMA observations showed the Keplerian nature of the rotation field \citep[e.g.][and reference therein]{Dutrey2014_PPVI}, it is only with the first ALMA observations that sufficient spatial resolution was reached to separate the emission originating from the upper and lower surfaces (\emph{i.e.}, above and below the midplane as seen from the observer),
and from the near and far sides of these surfaces \citep{Rosenfeld2013a,deGregorio-Monsalvo2013}.
These ALMA observations revealed details about the structure of the disks, they confirmed the vertical temperature gradients, the flaring shape of the molecular layer, and highlighted which limb is approaching and which limb is receding.

Because channel maps only integrate the received photons over a narrow bandwidth, they are usually noisy. An alternative way to look at kinematic data is to compute ``moment maps'' by integrating over the spectral axis. The most commonly used maps are the zeroth $M_0 = \sum_{i=1}^N I_i$ and first moment $M_0 = \sum_{i=1}^N I_i v_i/ M_0$, where $i$ denote the index of the velocity channel and $N$ the number of channels summed over. The zeroth moment shows the integrated intensity and reveals the overall morphology of the line emission, while the first moment shows the intensity weighted average velocity (see Fig.~\ref{fig:channel_maps}). These maps are also affected by the noise in the individual channels and often require threshold masks such that, in each pixel, a few channels with signal are not contaminated by a large number of channels dominated by noise. An additional potential issue of these moment maps is that at high spatial resolution, the upper and lower surface can be separated for some lines, such as $^{12}$CO (see Fig.~\ref{fig:channel_maps}, left). These two surfaces have different fluxes, making their average sometimes difficult to interpret. Alternative methods, for instance taking the peak intensity and velocity of the peak have been proposed to alleviate these issues. Slightly more advanced methods fitting a Gaussian profile \citep{Casassus2019b} or a quadratic function \citep{Teague2018d} around the peak intensity results in more accurate and less noisy maps. Using double Gaussian line profiles allows to approximately separate the velocity centroids of the upper and lower surfaces \citep{Casassus2021,Izquierdo2022}.

Assuming the gas orbital motions are dominated by Keplerian rotation, it is possible to ``invert'' the butterfly pattern using simple geometrical arguments, and to reconstruct the radius and altitude of each point along the isovelocity curve when the disk is seen at an intermediate inclination \citep{Pinte2018a,MAPS4}. Using multiple tracers, for instance a combination of CO isotopologues, such measurements allow for derivation of the velocities as a function of radius and altitude for an axially symmetric disk.

For optically thick lines, the observed brightness temperature $T_\mathrm{b} = T_\mathrm{ex} (1 - \exp(-\tau))$ is almost equal to the excitation temperature $T_\mathrm{ex}$ (as long as the emission fills the beam).
Because low J CO lines are close to LTE, the brightness temperature is also close to the kinetic temperature, and the peak intensity then provides an accurate estimate of the gas temperature at the $\tau \approx 1$ surface \citep[e.g.][]{Weaver2018}. In combination with the reconstructed positions, this allows to build a map of the disk temperature structure without the need to rely on models \citep{Pinte2018a,MAPS17}. A similar method was developed to reconstruct the altitude of the emitting layers in edge-on disks \citep{Dutrey2017,Teague2020,Ruiz-Rodriguez2021,Flores2021}.  When the two surfaces are separated, the lower surface can be used to measure the temperature of the gas just above the vertical snow-line \citep{Pinte2018a}, and the emission between the upper and lower surfaces can probe the midplane temperature \citep{Dullemond2020}, with the caveat that the emission can become optically thin and the brightness temperature become lower than the actual gas temperature.

The emission is centered on the iso-velocity curve corresponding to the central velocity of the channel. The spatial extent of the emission in the direction perpendicular to the iso-velocity curve is the results of 3 effects: the spectral resolution of the interferometer, the local line width, and the optical depth of the line. If the spectral resolution of the observations is high enough, it can be subtracted, or at least accounted for, to recover some information on the local emitting region. If the local line (thermal + turbulent) width is large (compared to the spectral resolution), emission from a given point in the disk will be received in multiple channels. Conversely, emission in a given channel will arise from a larger region of the disk than in the case of a narrow line width. If the line is highly optically thick, the core of the line will saturate and photons will escape from the wings of the line, \emph{i.e.} at high velocity from the core the line, and the emission will appear more extended in the direction orthogonal to the iso-velocity contour. This is particularly true for $^{12}$CO emission, when all the projected velocities ``align'', for instance near the semi-major axis of the disk at systemic velocity. Note that, even when the line is optically thin, or marginally optically thick, some photons will escape from the wings, but the limited dynamical range of the interferometer will prevent their detection.

Because the channel maps (and moment 1 or peak velocity maps) directly reflect the underlying gas velocity, they can be used to measure almost directly the Keplerian rotation of the disk, as well as deviations from the Keplerian profile. A perfectly Keplerian disk will result in smooth iso-velocity contours and velocity maps. Velocity perturbations will change the projected velocity of the emitting gas and shift the emission into an adjacent channel. In a given channel, the velocity perturbation will result in  ``kinks'' or ``wiggles'' modulating the butterfly pattern characteristic of Keplerian rotation.

\subsection{Rotation curves}
\label{sec:methods:rotation_curves}

The velocity field of the disk will be dominated by the (sub-)Keplerian rotation of the gas, as evidenced by the ``butterfly'' structure seen in the channel maps or the dipole morphology seen in the first-moment maps. There are several methods that have been adopted to extract the rotation curve from the data after making several assumptions which we describe below.

The most straight forward is the modeling of a first-moment map of an optically thick line, where the projection of the rotation velocity is given by $v_{\phi,\, {\rm proj}}(r,\, \phi) = v_{\phi}(r) \cos(\phi) \sin(i)$, with $\phi$ being the polar angle in the disk frame of reference and $i$ the inclination of the disk. Under the assumption of axial symmetry, and with knowledge of the orientation of the disk, i.e., its inclination, position angle, along with the source center and an independent knowledge of $z(r)$, the height of the emission surface, it is possible to transform on-sky positions to disk-frame coordinates \citep[see][for example]{Rosenfeld2013a}. By then adopting a model for $v_{\phi}$, for example a pure Keplerian profile as in Eqn.~\ref{eq:orbital_velocity_keplerian}, the free parameters of the model can be optimized to recover the best-fitting rotation curve. The Python package \texttt{eddy} \citep{eddy} provides this functionality. More recently, \citet{Izquierdo2021} presented the \texttt{discminer} package which performs a similar functionality, however does this by fitting the emission morphology in each channel, rather than only the velocity centroid.

Although a rapid approach to extracting the rotation curve, an analytical $v_{\phi}$ model will be unable to capture the subtle variations expected in the rotation curve owing to changes in the emission height, the radial pressure gradient and the additional gravitational potential of the disk described in Eqn~\ref{eq:orbital_velocity_total}. To circumvent this issue, several works have opted to fit concentric annuli independently in order to recover a non-parametric model of $v_{\phi}(r)$, for example as performed by the \texttt{ConeRot} package \citep{Casassus2019}. The advantage of these approaches is the flexibility that they offer, however are far more susceptible to noise.

With the more flexible `independent annuli' approach there are have been two main methods employed to infer the correct velocity. \citet{Casassus2019, Casassus2021} consider the line centroids, which results in an extremely fast calculation, however requires some thought about the most appropriate way to measure the centroids (see the discussion in Sec.~\ref{sec:channel_maps}). An alternative approach has been presented by \citet[and also available with \texttt{eddy}]{Teague2018a, Teague2018c}, where the velocity structure is inferred by searching for the $v_{\phi}$ at a given radius that best aligns all spectra when they are corrected for the projected component of $v_{\phi}$. \citet{Yen2018} applied such a technique to constrain the dynamical mass of 41 young stars in the Lupus region. These approaches are the inverse of a commonly used technique to extract weak molecular line signals from observations of protoplanetary disks where a velocity structure is assumed such that all emission can be aligned and thus stacked \citep[e.g.,][]{Yen2016,MAPS9}. A similar method is routinely used in the detection of molecules in the atmospheres of exoplanets by leveraging the known orbital frequency of the planet \citep[and thus time-dependence of the Doppler shift;][]{Snellen2010}.

The fundamental assumption with all these methods is that the underlying velocity structure of the disk is azimuthally symmetric, that is, the only $\phi$ dependence in $v_{\phi,\, {\rm proj}}$ is from the $\cos(\phi)$ projection term. However, these azimuthally symmetric background models provide an opportunity for localized deviations to be identified, enabling additional iterations of inference excluding these regions. Additionally, with ALMA routinely achieving spatial resolutions necessary to disentangle the upper and lower surfaces of the disk, emission from the lower surface of the disk may contaminate the top side when the emission is optically thin \citep[e.g.][]{MAPS18}. However, both \citet{Casassus2021} and \citet{Izquierdo2021} demonstrated that by fitting two Gaussian components the front and back sides of the disk can be disentangled and thus analyzed separately.

\subsection{Limitations}
\label{sec:limitations}

Unlike detecting emission from a source where collapsing of the data cube can be leveraged to improve the signal-to-noise to improve the detection, extracting the velocity structure requires characterizing the emission morphology in the full 3 dimensional position-position-velocity space native to sub-mm interferometers. This presents several challenges, both in the collection and analysis of the data.

From a data collection standpoint, programs need to be designed to fall in the expensive high angular and spectral resolution regime. While observations of dust continuum can leverage the full long baselines of facilities like ALMA, they are only able to do so by averaging data over large bandwidths (${\sim}$~2~GHz). With the necessity of having to spectrally resolve line emission, and typical FWHM of ${\approx}~200~{\rm m\,s^{-1}}$, kinematic observations require spectral resolutions at close to the limit of 31~kHz (${\approx 40~{\rm m\,s^{-1}}}$ at 230.538~GHz). Due to the spectral response of the correlator \citep{ALMA_technical_handbook}, this is particularly important when looking at the broadening of lines as the resulting spectra will be artificially broadened, potentially masking the true level of broadening.

A further complication when working with interferometric images of molecular line data, particularly when searching for subtle deviations in the emission morphology indicative of localized velocity perturbations, is that the imaging process (the conversion of the recorded Fourier visibilities to a 3D image and the subsequent \texttt{CLEAN}ing of the data to remove substantial side-lobe features, e.g. \citealp{Hogbom1974}) can introduce artefacts that may produce false-positive detections. To circumvent these issues, or at least provide a secondary point of reference, a family of new imaging techniques broadly referred to as `regularized maximum likelihood' (RML) techniques are becoming commonplace \citep{Cornwell1985,Narayan1986,EHT2019}. Although beyond the scope of this chapter, the importance of the development of these new techniques (\texttt{GPUVMEM}: \citealp{Carcamo2018}, \texttt{MPoL}: \citealp{mpol}) will substantially benefit future studies of gas kinematics.

As pointed out by \cite{Boehler2017} and \cite{Weaver2018}, caution is also required when interpreting line emission in regions where there is continuum emission. Optically thick dust can absorb the line (depending on the relative position of their emitting regions), distorting the line emission and impacting any kinematic analysis.
If the line is optically thick, another effect can affect the analysis. The line will absorb the dust continuum. Because the continuum emission is usually estimated by interpolating the emission between line-free channels, it will be overestimated near the line center. By subtracting the overestimated continuum, the line flux will be underestimated. Depending on the respective morphologies of continuum and line emission, this can create distortions in channel maps that mimic velocity deviations. Without knowing the relative optical depths of the continuum and line, it is impossible to correct for these effects. Imaging with and without continuum subtraction can help assessing if they are impacting the data.

An additional complication is the impact of spatial resolution on the characterization of the disk velocity structure.
Beam smearing biases velocities when steep line intensity gradients are present \citep{Keppler2019,Boehler2021}. Inside a beam, the spectrum, and thus the measured velocity of the gas, are an average of all the spectra sampled by the beam. If the intensity gradient is steep across the beam, this will bias the velocity towards the region of high intensities, rather than the center of the beam. As there are very large velocity gradients close to the host star, even observations with moderate beam sizes will smear out the emission leading to highly non-Gaussian line profiles with long tails. The effect is even more complex if local intensity gradients are present. In the case of a gap, this leads to artificial super-Keplerian velocities at its inner edge, and sub-Keplerian velocities at its outer edge. This effect is opposite to the actual signature expected from the pressure gradient, but the measured deviations from Keplerian rotation are a combination of the two (see Appendix A.2 in \citealp{Keppler2019}).  \cite{Boehler2021} further demonstrated that this beam smearing effect can mimic anti-cyclonic vortices near azimuthal asymmetries. This effect cannot be properly accounted for without knowing the underlying intensity profile.
Although performing forward modelling in the $uv$-plane will aid here, only longer-baseline observations (resulting in smaller beam sizes) will allow us to probe the velocity structure of the inner regions of disks, or near steep line intensity gradients. Combining observations from multiple lines with different optical depths (and hence different brightness gradients) can also help to mitigate these effects, but with the added complexity that the various lines will sample a range of altitudes.

\begin{figure*}
  \includegraphics[width=\hsize]{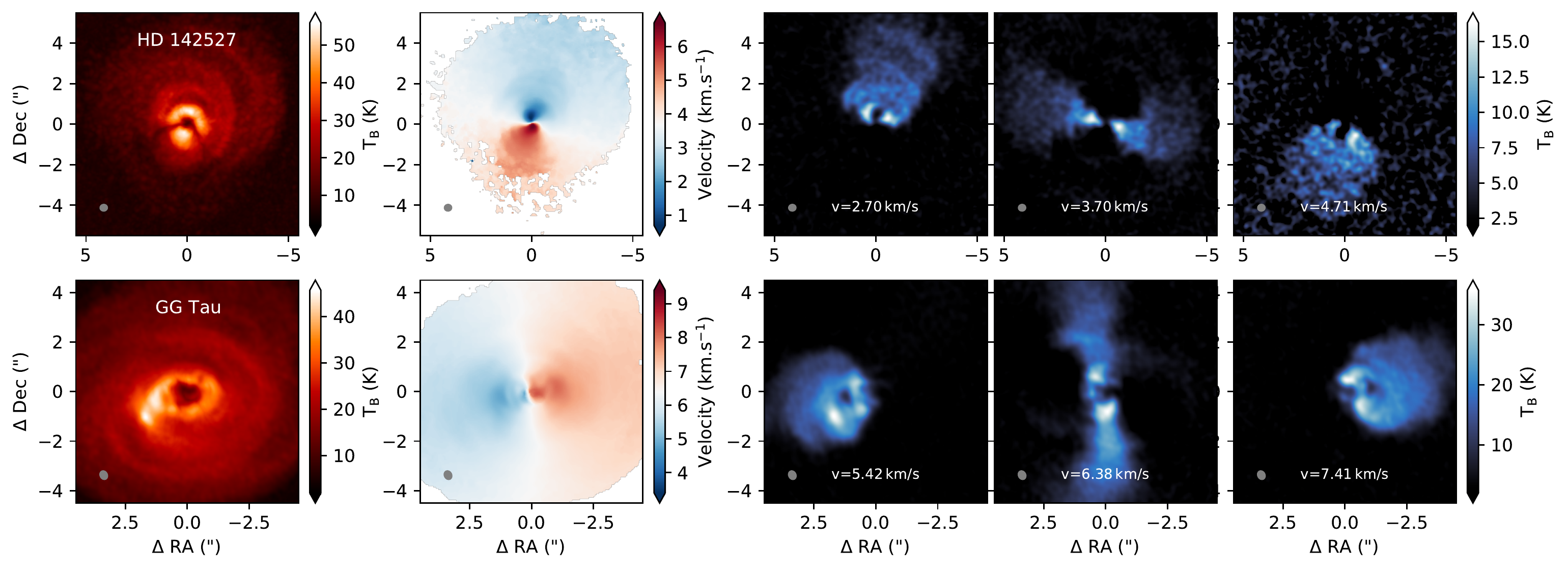}
  \caption{Examples of non-Keplerian motions in $^{12}$CO J=2-1 emission of the circumbinary disks of HD~142527 \citep{Casassus2013,Christiaens2014,Garg2021} and GG~Tau \citep{Phuong2020}.  Moments 0 (left) and 1 (center), and selected channel maps (right). Note that the dark V-shape is the integrated intensity map of HD~142527 is an artefact due to foreground cloud extinction over part of the line. Both objects display a central cavity carved by the multiple systems, as as large scale spirals. The spirals are also detected in the outer parts of the moment 1 maps, while the central parts show "twisted" kinematics indicative or fast radial flows and/or tilted inner disks. The same kinematic structures manifest as multiple distortions over the whole disk and at all velocities.\label{fig:TD}}
\end{figure*}

A final limitation is the optical depth of the selected lines. All the methods described earlier rely on the assumption that the emission arises from an infinitely thin layer in the disk. For optically thick molecules, like $^{12}$CO, this approximation is mainly valid, and kinematic structures are clear. But the high optical depth of $^{12}$CO means that there is often cloud contamination (either foreground extinction and/or spatial filtering of the interferometer if large scale emission is present). Disk emission can be decreased over some channels, in practice hiding part of the disk and distorting the moment maps (see for instance top left panel in Fig.~\ref{fig:TD} for HD~142527). If the contamination is not too strong, it can be corrected for to allow kinematic studies  \citep[e.g.][]{Garg2021}. But in most cases, the contamination is so strong that some channels have no recoverable emission, which impacts the ability to understand the nature of some structure.
For instance, some of $^{12}$CO J=2-1 observations in the DSHARP sample are heavily contaminated, hiding parts of the disks. As a result, \cite{Pinte2020} were not able to determine if the tentative kinematic detections are localized or are present across the whole disk.
One solution to limit this contamination is to move to higher transitions, like J=3-2 or J=6-5 instead of J=2-1, as the surrounding cloud is mostly cold ($\lesssim$ 15K) and becomes less optically thick at higher frequencies.
Another option is to move to less abundant molecules, like $^{13}$CO or C$^{18}$O, but the lower optical depth means that the emission is coming from a thicker layer in the disk, resulting in projection effects in both the plane of the disk and the vertical direction, making the process of reconstructing the velocities far more complex.

When it comes to interpreting these data, a common approach is to model the data with some parametric form \citep[e.g.,][]{Rosenfeld2013a, Williams2014}. However, with the quality of current observations it is clear that the complexity of these systems far exceeds that which is easily parameterized. Disentangling subtle variations in the emission morphology therefore requires a suitably flexible, but simultaneously well-motivated, model. Future observations that better characterize the physical structure of these sources  will be essential in developing such intuition and allowing users to tease out even smaller velocity substructures.

\section{\textbf{Large-scale gas motions in disks}}

\subsection{Dynamical mass measurements}
\label{sec:dynamical_mass}

Precise measurements of the physical properties of pre-main sequence stars is fundamental to understand their evolutionary status, build initial stellar mass functions and set the timescale for disk evolution and planet formation. Eclipsing binaries have been the most robust method to test the theoretical predictions of stellar evolution models \citep[e.g.][]{Stassun2014}.
Spatially and spectrally resolved ALMA data allows for comparison with detailed radiative transfer disk models and leads to independent measurements of the central star mass, with typical uncertainties $\lesssim 5\%$  \citep{Rosenfeld2013b,Czekala2015,Czekala2016,Czekala2017,Sheehan2019,Simon2019,Pegues2021}, \emph{i.e.} a factor a few lower than in the pre-ALMA area  \citep[e.g.][]{Simon2000}.
An intrinsic limitation of any dynamical estimate is the ambiguity with the inclination. \cite{Casassus2021} showed that the derived stellar mass depends on the inclination $i$ as
\begin{equation}
M_* \propto r \times v_\phi^2 \propto \frac{1}{\cos i} \times \frac{1}{\sin^2 i}.
\end{equation}
All the terms discussed in sec.~\ref{sec:velocity_fields} also impact any dynamical mass estimate, and care must be taken when analyzing the data, especially when the disk is only marginally spatially resolved.

\cite{Yen2018} and \cite{Braun2021} used the technique of velocity-aligned line stacking to increase signal-to-noise and infer dynamical masses for fainter objects in the Lupus and Taurus star forming regions. The comparison of these dynamical masses with spectroscopic estimates show that none of the current stellar evolution models can provide a robust mass estimate over a wide range of masses from 0.1 to 1.4\,M$_\odot$, confirming the seminal results from \cite{Hillenbrand2004}. Magnetic models \citep{Feiden2016} provide the best estimates for masses above 0.6\,M$_\odot$, while non-magnetic models \citep{Siess2000,Baraffe2015,Feiden2016} appear more accurate for lower masses.

Using gas kinematics to infer stellar masses can also clarify the nature of faint objects. For instance, ALMA CO observations of
the planet candidate FW Tau C, initially detected via direct imaging \citep{White2001}, revealed that it is unlikely a planet, but rather a low-mass star with a highly inclined disk \citep{Wu2017}.

Disks in Keplerian rotation are also commonly detected around Class I protostars \citep[e.g.][]{Brinch2007,Takakuwa2012,Lee2016}.
In a similar way as for class II objects, comparison with radiative transfer models has been the method of choice to measure dynamical masses.

However, disks in young Class 0 protostars have
largely remained elusive up to now, as it is intrinsically difficult to isolate a disk in envelope-dominated objects \citep[e.g.][]{Andre1993}. The detection of protostellar outflows indicates the presence of disks at the earliest stages of the protostar evolution \citep{Arce2007}. The detection and characterisation of these disks at early stage is central to understand their formation and the initial conditions for the building of planets.
Recent surveys of Class II disks show that most disks are smaller than 30\,au (e.g. \citealp{Ansdell2016}, see chapter by Miotello et al.), suggesting they might be even smaller at earliest stages if they are viscously spreading. MHD simulations of protostellar disk formation also suggest that disks do not extend over more than about 100 au in size \citep[e.g.]{Seifried2012,Joos2013,Myers2013,Hennebelle2016,Hennebelle2020}. Only high spatial resolution ($\lesssim$ 0.1'') sub-millimetre observations can shed led light  on the characteristics of protostellar disks \citep{Seifried2016}.
As such, a large effort has been made in recent years to observe the dynamics of disks around Class 0 objects. High spatial resolution sub-millimeter observations have revealed direct evidence of Keplerian rotation, but only for a few objects \citep{Tobin2012,Hara2013,Murillo2013,Yen2013,Sanchez-Monge2013,Codella2014,Lee2014,Ohashi2014,Tobin2016,Aso2017,Yen2017a,Reynolds2021}. Other observing programs have led to non-detection \citep{Maury2010,Yen2013,Yen2015,Maret2014}.
A survey of 16 Class 0 sources by \cite{Maret2020} found evidence of Keplerian rotation in only two sources, concluding that Keplerian disks larger than 50\,au are rare around Class 0 sources.
The non-detection on smaller scales could be a spatial resolution issue, and it is still unclear how common Keplerian disks are around Class 0 protostars (see chapter by Pineda et al. for a more detailed discussion.)

\subsection{Constraints on turbulence and implications on angular momentum transport }
\label{sec:turbulence}
The main physical ingredient that drives mass accretion and angular momentum transport or removal in disks remains unknown. Whether it is turbulence or some other process is critical for a range of planet formation stages, as it affects the redistribution and growth of dust grains, the delivery of planet building material to newly formed cores, and the subsequent migration of embedded protoplanets. Observational constraints on the turbulence, as well as identification of the process at play, is essential if we are to decipher the planet formation process.

\begin{figure}[!t]
  \includegraphics[scale=1.0]{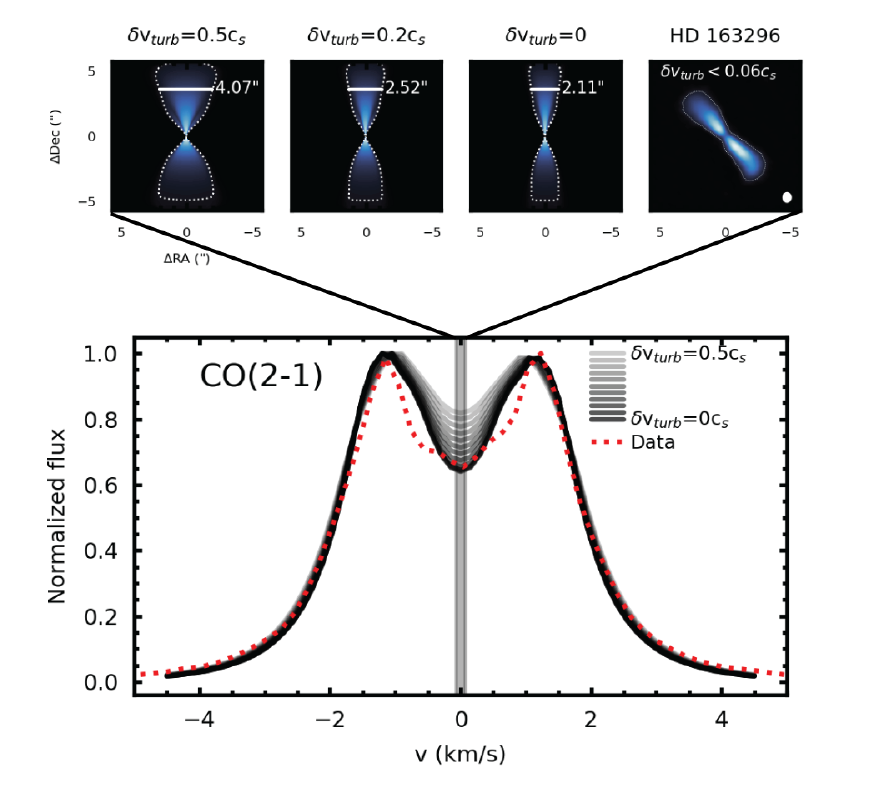}
  \caption{Turbulence generates distinct observational features, producing substantial broadening in the individual channels, as well as a change in the line peak-to-trough ratio, as seen in the HD~163296-like CO(2-1) model disk emission ($i=48^{\circ}$, M$_{*}=2.3M_{\odot}$, M$_{\rm disk}=0.09M_{\odot}$, \citealt{Flaherty15}) shown above. The effect of turbulence is distinct from that of gas temperature, which more strongly affects the strength of the emission for these optically thick lines. Note that turbulence still imprints an observational signature on the data, even though for $\delta v_{\rm turb}<$0.4c$_s$ the turbulent motion in the outer disk is below the velocity resolution of this model (150~${\rm m\,s}^{-1}$).
  \label{fig:turbulence}}
\end{figure}

Resolved molecular line observations provide a direct measure of the turbulent motion, and ALMA observations at high spectral and spatial resolution have resulted in a paradigm shift in our understanding of the accretion processes in disks. While pre-ALMA studies \citep{Hughes2011,Guilloteau2012} were limited in both spectral resolution and sensitivity, ALMA is able to obtain high signal-to-noise data even at high spatial resolution. At the modest turbulent velocities expected in planet-forming disks, the effect of turbulence is most clearly seen as a broadening in the channel maps and a change in the peak-to-trough ratio of the spectrum \citep[Fig.~\ref{fig:turbulence},][]{Simon2015}. This is because the turbulence offsets some of the orbital velocity due to Keplerian motion, allowing gas emission to leak in from neighboring channels. Temperature also contributes to line broadening, but this can be disentangled from turbulence, for instance by  using multiple line transitions to constrain the gas temperature \citep{Teague2018c}, or taking into account of the effect of temperature on the intensity of optically thick emission \citep[e.g.][]{Flaherty2020}. The latter effect is most strongly dependent on the amplitude calibration uncertainty that, at $\sim$10\% \citep{Butler2012}, is often the largest source of uncertainty in ALMA observations. Taking into account these effects has produced upper limits of 5-10\% of the sound speed (at radii larger than $\approx$ 30-100 au, \citealp{Flaherty2015,Teague2016,Flaherty2017,Flaherty2018,Teague2018c}), with a detection in DM Tau at 0.25 c$_s$ to 0.33 c$_s$ \citep{Flaherty2020}. As a zeroth-order assumption, these studies use a single turbulence value for the entire disk, although turbulence is expected to vary across the disk (see Lesur et al. chapter in this volume) and potential evidence for spatially varying non-thermal line-widths has been seen in the spiral arms around HD~135344~B \citep[$\sim0.22c_s$,][]{Casassus2021} and HD~142527 \citep{Garg2021}, and in clumps around RU Lup \citep{Huang2020b}.

In the context of the $\alpha$-disk model of viscosity \citep[$\nu = \alpha c_s H$,][]{ShakuraSunyaev1973}, turbulence can be related to $\alpha$ via $\alpha\sim(\delta v_{\rm turb}/c_s)^2$ (see Lesur et al. chapter in this volume for a more detailed discussion of $\alpha$ and its relation to turbulent velocities). The turbulence measurements imply $\alpha\lesssim$ a few$\times10^{-3}$, with $\alpha\sim$0.06-0.11 for DM Tau. The low $\alpha$ values are broadly consistent with measurements in other systems, which typically find $\alpha = 10^{-4} - 10^{-3}$, based on the vertical/radial diffusion of dust grains \citep{Mulders2012,Pinte16,Boneberg2016,Dullemond2018,Rosotti2020,Ohashi2019,Ueda2021a,Doi2021}, the size of disks \citep{Najita2018,Trapman2020}, and the relationship between the accretion rate, size and mass of disks \citep{Ansdell2018,Ribas2020}. Despite the similarity in these results, caution must be used in directly comparing these different observations. Measurements of $\alpha$ based on the dust scale height depend on the grain size, which may differ between observational diagnostics \citep{Ueda2021a}. The dust diffusivity may vary between the radial and vertical direction leading to different $\alpha$ values derived from the vertical or radial dust scale lengths, which can have important implications on the grain growth process \citep{Pinilla2021}. Constraints on $\alpha$ from planet-induced gap depths \citep[e.g.][]{Liu2018} depend strongly on the assumed planet mass and disk scale height \citep[e.g.][]{Kanagawa2015}. Comparisons of observational results also depend on the vertical (and radial) region being probed. Dust settling and ring radial widths are most sensitive to turbulence at the midplane \citep{Ciesla2007}, while optically thick CO emission is sensitive to turbulence in the upper layers of the disk. In the context of magneto-rotational instability (MRI) and vertical-shear instability (VSI), the turbulent velocity is expected to decrease towards the midplane, leading to smaller $\alpha$ values at the midplane compared to the upper layers of the disk. For a given turbulent velocity, VSI is more efficient at lofting dust grains into the disk atmosphere than MRI \citep{Flock2017,Flock2020}, affecting the comparison of velocity and dust scale height measurements.

Despite these limitations, the turbulence velocity constraints can be directly compared to predictions from simulations to better understand angular momentum transport in disks. In the context of the magneto-rotational instability with sufficient ionization and magnetic field strength, the velocities in the high molecular layers where much of the optically thick emission arises are expected to be 10-100\%\ c$_s$ \citep{Simon2018}. These velocities depend strongly on the ionization, which has been measured in a handful of disks \citep{Cleeves2015a,Seifert2021} and the magnetic field strength, which can be measured in our solar system \citep{Weiss2021}, but only upper limits are as yet available in other systems \citep{Vlemmings2019,Harrison2021}. In the absence of MRI, hydrodynamic instabilities may operate (see review by \citealp{Lyra2019}). While gravito-turbulence \citep{Forgan2012,Shi2014,Shi2016} predicts velocities comparable to that of MRI, and is therefore ruled out for most of the observed systems, the VSI \citep{Flock2017,Flock2020} predicts velocities $\sim$0.05c$_s$, closer to the upper limits.

\subsection{Evidence for winds as a main ingredient for angular momentum removal}
\label{sec:wind}

Without strong evidence for MRI-active disks, disk winds may be the main driver of angular momentum removal \citep{Bai2016}. Numerical models indicate that disk winds generate turbulence near the base of the wind \citep{Cui2021}, at levels consistent with the upper limits from molecular line observations. The bulk motions generated by disk winds produce their own distinct kinematic signatures, as discussed in Pascucci et al. (this volume), which provide a more direct measure of this process \citep{Blandford82, Konigl91, Pudritz06PPV}.

Searches for these signatures of disk winds have primarily focused on the inner regions of the disk employing optical spectroscopic measurements to identify high- and low-velocity components at small radii \citep[e.g.,][]{Banzatti2019, Pascucci2020}. However, with access to the kinematics of the outer regions of the disk, signs of outflows capable of transporting angular momentum are increasingly common for Class 0/I sources, with CO, CS and SO observations tracing winds launched from radii between ${\sim}~1$ and ${\sim}~40$~au \citep{Launhardt2009, Bjerkeli2016, Tabone2017, Zhang2018_wind, Zhang2019_wind, Hirota2017, deValon2020}.

Evidence for winds in Class II sources is much scarcer, with only two sources displaying evidence for winds: HH-30 \citep{Louvet2018} and HD~163296 \citep{Klaassen2013}. For the edge-on HH-30, the outflow is only observed on one side of the disk and found to be launched from within 22~au and extending out to over 700~au. Measuring a wind velocity of $9.3 \pm 0.7~{\rm km\,s^{-1}}$, \citet{Louvet2018} place a minimum mass flux of $9 \times 10^{-9}~M_{\odot}~{\rm yr}^{-1}$, consistent with the predictions of winds launched through photo-evaporation or magneto-centrifugal processes. HD~163296 hosts a large scale wind or outflow, originally detected through $^{12}$CO emission associated with several HH-knots. \citet{Klaassen2013} measure a wind velocity of $18.9 \pm 0.5~{\rm km\,s^{-1}}$, resulting in an estimated mass removal rate of $(6.2 \pm 0.7) \times 10^{-9}~M_{\odot}~{\rm yr}^{-1}$. Intriguingly this outflow rate results in a mass loss rate almost twice the accretion rate measured by \citet{Garcia-Lopez06}. \citet{Teague2019} observed a persistent radial outflow along the $^{12}$CO emission surface in the outer regions of the disk which the authors argued could be the base of such a wind. More recent observations of the wind by \cite{MAPS16} trace it in lower energy $^{12}$CO transitions and $^{13}$CO emission, enabling a better constraint on the excitation conditions of the wind. It was found that the wind was cooler than previously expected, resulting in a denser wind and thus a large mass loss rate of $4.8 \times 10^{-6}~M_{\odot}~{\rm yr}^{-1}$, resulting in an ejection-to-accretion ratio of between 5 and 50.

\subsection{Pressure profile, velocity gradients and condition for hydrodynamical instabilities}
\label{sec:pressure}

\begin{figure*}[t]
  \includegraphics[width=\textwidth]{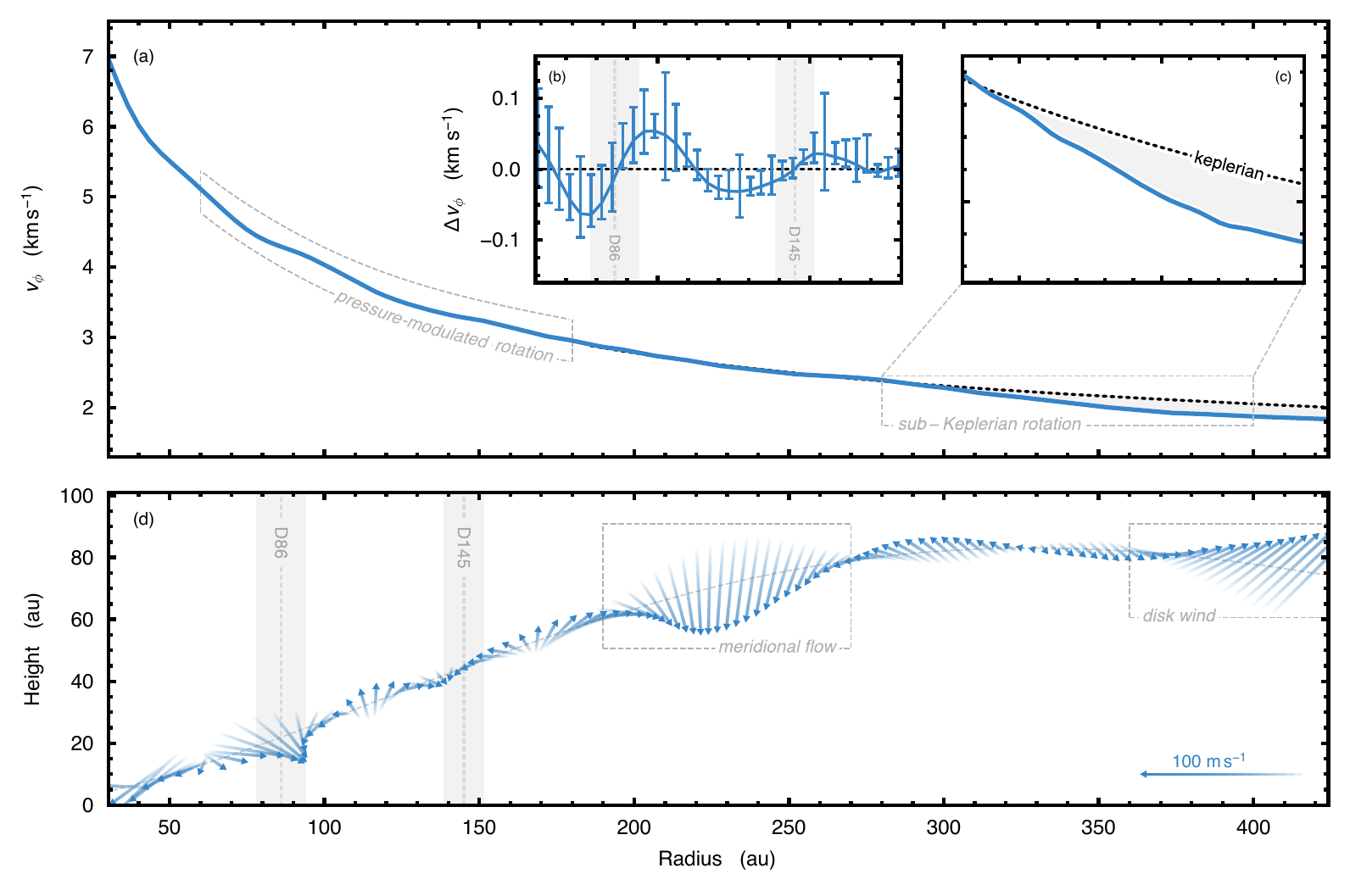}
  \caption{(a) Rotation curve derived from $^{12}$CO observations of HD 163926 \citep{Teague2019,MAPS18}. The disk mostly rotates at Keplerian velocities, except in the outer regions, likely an effect of the negative pressure gradient as the gas density drops in the outer disk, highlighted in panel (c). In inner regions, gaseous substructures cause the hastening and slowing of the gas rotation, highlighted in panel (b) which shows the Keplerian-subtracted $v_{\phi}$ value. (d) The inferred $v_r$ and $v_z$ components following the method presented in \citet{Teague2019}. \label{fig:rotation_curve}}
\end{figure*}

By leveraging the information from individual channel maps at high spectral resolution, \cite{Pinte2018a} developed a method to locate the CO isotopologue emitting surfaces, allowing the direct measurements of radial and vertical temperature and velocity gradients. They showed that the orbital velocities become sub-Keplerian where the surface density profile transitions from a power-law to a tapered structure, at $\approx 300$\,au, \emph{i.e.}, where the pressure gradient becomes steeper as expected from equation~\ref{eq:pressure_gradient}. \cite{Dullemond2020} found a similar tentative result for HD~163296, where the shape of the $^{12}$CO  emission in low-velocity channels (including emission from the outer parts of the disk) is better reproduced by a sub-Keplerian disk. The transition between Keplerian and sub-Keplerian seems to occur at $\approx$ 300\,au, \emph{i.e.} again where the tapered structure started \citep[e.g.][]{Isella2018}. Interestingly, \cite{Dullemond2020} found that the surface density in the outer disk needs to drop faster than the typical $\exp(-r/r_c)$ profile from viscous evolution \citep{LyndenBell74}, \emph{i.e.} corresponding to a steeper negative pressure gradient to account for the observed sub-Keplerian flow.
For IM Lupi, the radius at which velocities become sub-Keplerian also corresponds to the outer edge of the dust disk, consistent with the fact that dust migrates faster when the gas becomes more sub-Keplerian (as the velocity differential, and hence gas drag, increase between the dust and gas phases), while for HD~163296 the dust is much more compact. Fig.~\ref{fig:rotation_curve} presents an extraction of the velocity curve for the $^{12}$CO emission of HD~163296, confirming the results by \cite{Dullemond2020}.

By reconstructing the azimuthal velocities at different altitudes from a combination of CO isotopologues, \cite{Pinte2018a} also revealed vertical gradients of azimuthal velocities where the upper layers of the disk, probed by $^{12}$CO rotate slower than lower layers, for instance probed by $^{13}$CO and C$^{18}$O. The differences between the $^{12}$CO and $^{13}$CO layers are consistent with the vertical dependence of the stellar gravity (Eq.~\ref{eq:velocity_vertical_dependence}). The variation of orbital velocities with altitude also provide some additional information on the thermal structure. In particular, this confirms that the disk has a baroclinic structure, as the vertical gradient of velocity violates the Poincar\'e-Wavre theorem (\citealp{Wavre1929}, see also \citealp{Tassoul2000}, section 3.2.1 and \citealp{Stahler2004}, section 9.2.1 for discussions of this theorem on stellar interior and molecular clouds respectively). This vertical gradient of the azimuthal flow is equivalent to one driven by the baroclinic structure in a planet atmosphere also known as the (somehow misnamed) "thermal wind".
These vertical velocity and pressure gradients also provide observational evidence that the required conditions for hydrodynamical instabilities (such as the VSI, \citealp{Nelson2013,McNally2015}) are met in protoplanetary disks. As mentioned in the previous section,  these hydrodynamic instabilities likely contribute to some additional angular momentum transport.

Using high angular resolution observations of HD~163296, AS~209, and HD~169142, \citet{Teague2018a, Teague2018c, Yu2021} were able to use radially resolved variations in $v_{\phi}$ to infer the presence of localized variations in the radial gas pressure structure. Using forward models to match the data, they were able to show that these variations were likely caused by depletions -- gaps -- in the gas surface density profile, resulting in a characteristic hastening and slowing of $v_{\phi}$ along the inner and outer edge of the gap, respectively \citep[e.g.,][]{Kanagawa2015}. A substantial hurdle in such an analysis is removing an accurate model of the Keplerian component of the rotation, \emph{i.e.}, the first term on the right hand side of Eqn.~\ref{eq:pressure_gradient}, requiring a robust measurement of the emission height and the dynamical stellar mass \citep[e.g.,][]{Teague2019}.

With a measurement of the radial pressure gradient now accessible, it is possible to test the grain-trapping scenario proposed by \citet{Whipple1972} as a solution to the rapid radial drift of large grains. \citet{Dullemond2018} showed that as rings in the sub-mm continuum could now be spatially resolved, it was possible to place broad constraints on the trapping efficiency depending on the grain size distribution. Building on this, \citet{Rosotti2020} used the DSHARP $^{12}$CO emission to measure $v_{\phi}$ across the rings studied in \citet{Dullemond2018} to infer the profile of the pressure profile. Confirming that the pressure maxima were broader than the continuum rings demonstrated that these were indeed pressure traps, with the ratio of widths suggesting a strong level of coupling between the gas and dust.

\subsection{Large scale spirals and gravitational instability}
\label{sec:GI}

Large-scale spiral structure has been observed in protoplanetary disks at both micron \citep{Benisty2015,Stolker2016} and mm wavelengths \citep{Perez2016,Dong2018,Huang2018c}, suggesting that it is present at both large scale heights and the disk midplane. Unlike ring morphology, which the field is now generally reaching a consensus is due to planets, the origin of these spiral features has remained unclear.

One possibility is that they may be due to gravitational instability (GI). The degree to which a disk is self-gravitating is largely governed by the Toomre $Q$ parameter \citep{Toomre1964}, given by:
\begin{equation}
 Q=\frac{c_{s} \kappa}{\pi \mathrm{G} \Sigma}
 \label{eq:toomre}
\end{equation}
where $c_{s}$ is sound speed, $\kappa$ is epicyclic frequency, G is the gravitational constant and $\Sigma$ is the disk surface density. Essentially, pressure ($c_{\rm s}$) and rotation ($\kappa$) act against gravity ($\Sigma$), so that for $Q\gtrsim 2$ the disk is stable to GI \citep{durisen2007}.

Historically, the difficulty in determining the origin of spirals has come from the morphological similarity between the various scenarii. For example, in the MWC 758 system, planets \citep{Dong2015} and GI \citep{Dong2015GI} are both a viable explanation for the large-scale spiral structure observed. On the other hand, GI spirals are expected to take a logarithmic form with constant pitch angles of $10^\circ \lesssim \phi_{\rm{pitch}} \lesssim 15^\circ$ \citep{Cossins2009,Hall2016,Forgan2018,Forgan2018tache} while planet induced spirals are better fit by a radially varying pitch angle \citep{Goodman2001,Rafikov2002,Muto12,Zhudong2015}. Discerning between these scenarios requires high spatial resolution, and may be complicated by uncertainties introduced by deprojection effects, the region of parameter space under consideration, and low contrast ratios between arm and inter-arm regions \citep{Meru2017,Hall2018,Hall2019}.

\begin{figure*}[!t]
\includegraphics[width=\textwidth]{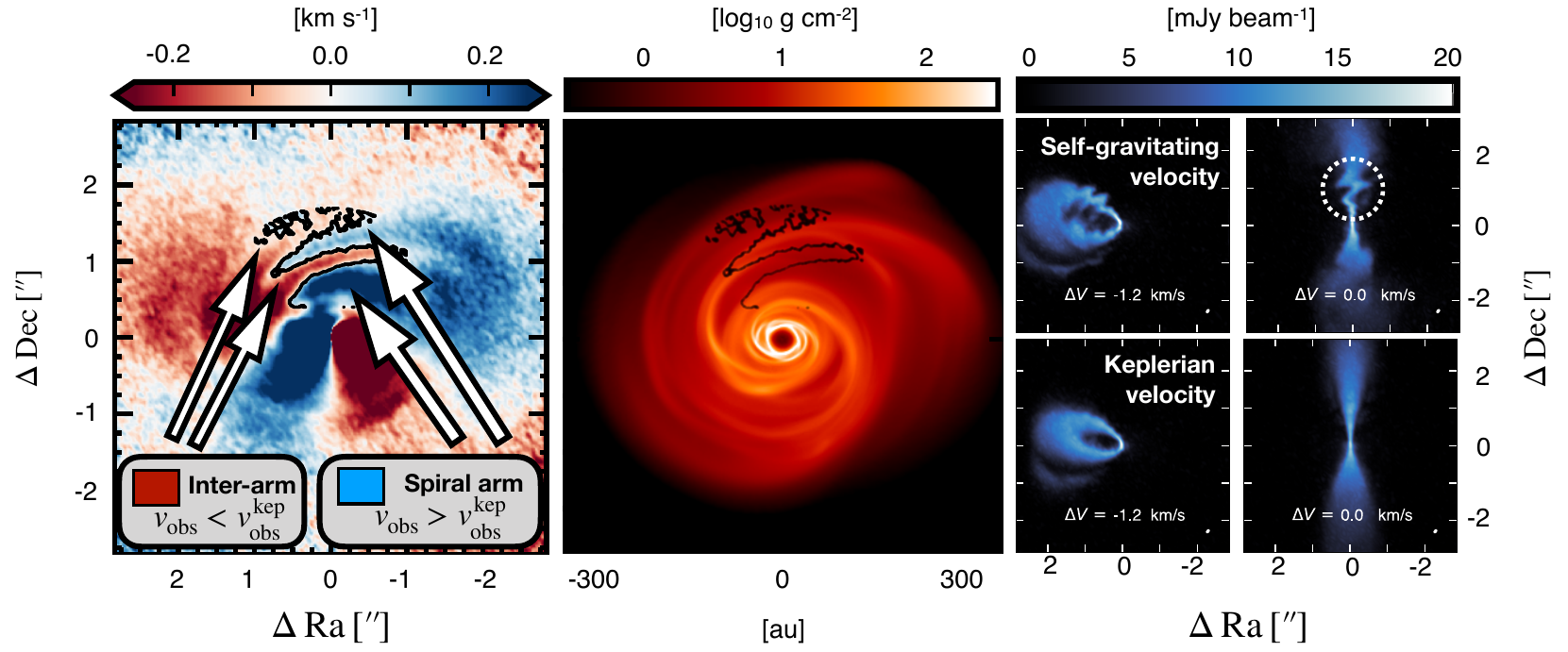}
\caption{Gravitational instability causes distinct kinematic features that are readily observable. Leftmost panel shows line-of-sight velocities with Keplerian rotation subtracted, resulting in ``interlocking finger'' structure. The line $v_{\rm{obs}} - v_{\rm{obs}}^{\rm{Kep}} =0  $ is shown in black. Center panel shows surface density at inclination of 30$^{\circ}$. Right panels show selected channels in the $^{13}$CO $J=3\rightarrow 2$ transition (top), compared to the same model but with the velocities set to Keplerian (bottom). The GI wiggle is circled at the systemic velocity. Images adapted from \citet{Hall2020}. \label{fig:wiggle}}
\end{figure*}

It has recently been demonstrated that kinematics may hold the key to conclusively determining the presence of GI. \citet{Hall2020} demonstrated that a disk undergoing GI should have two kinematic features. The first is the presence of the ``GI Wiggle'' - a distinctive zig-zag shape observed in line emission (rightmost panel of Figure \ref{fig:wiggle}), that is present across the azimuth of the disk and robust to viewing angle. Unlike spirals caused by embedded planets, where the velocity deviation from the azimuthal background is ultimately sheared out by the disk viscosity, GI spirals globally and continually perturb the disk velocity. Inside the spiral arms, velocity relative to the average background velocity is increased. This results in emission being pushed into the adjacent channel corresponding to a faster velocity along the line of sight to the observer. The inter-arm regions, on the other hand, have the the opposite effect. This results in the second kinematic feature of GI, which is the ``interlocking finger'' structure shown in the leftmost panel of Figure \ref{fig:wiggle}.

The GI-Wiggle, being a wave-like feature, is characterised by both wavelength and amplitude. \citet{Terry2022}  demonstrate that under the assumption of similar cooling properties between disks, it will be possible to constrain disk mass through characterisation of the GI wiggle in observations due to a positive relationship between wiggle amplitude and disk mass. A full analytical description of the GI Wiggle is given in \citet{Longarini2021}, who find that the shape of the wiggle depends disc mass, cooling, spiral pitch angle and the number of spiral arms.

The most promising system for the conclusive detection of active GI is Elias 2-27 \citep{Perez2016}. Indeed, recent observations and analysis by \citet{Paneque2021} have shown kinematic evidence for GI in the form of the GI Wiggle as described in \citet{Hall2020}, in addition to evidence of localized dust trapping consistent with GI spirals \citep{Rice2004,Gibbons2012,Gibbons2014,Cadman2020}.

The wiggles in Elias 2-27 are co-located with the spiral features, are stronger than typical of planetary-mass perturbers and are also present across a wide range of velocity channels. However, the spatial resolution at $\sim 0.2$ arcsec and the spectral resolution at 0.1 km/s was a factor of 2 and 3 respectively lower than the values used in \citet{Hall2020}, so higher spectral and spatial resolution observations are required to definitively confirm the presence of the GI Wiggle in the Elias 2-27 system. \citep{Terry2022} demonstrated that given sufficient spatial and spectral resolution, this is possible for a variety of disk masses. Interestingly, the presence of a planet can suppress GI by driving an increase in the Toomre parameter \citep{Rowther2020b}. If higher spectral and spatial resolution observations of GI reveal GI Wiggles with smaller amplitude than expected for the disk mass of Elias 2-27, then a planet in the prominent central gap \citep{Perez2016, Andrews2018} may be to blame.

Another promising avenue for determining the presence of GI is the dynamical measurement of disk mass from rotation curves. GI is typically active for disk-to-star mass ratios, $q$, of $\sim$ 0.1 or greater \citep{Kratter2016}. Constraining the disk mass therefore also places constraints on the likelihood of GI to currently be active.

The effect of self-gravity on the disk rotation curve is a promising avenue for measuring disk mass. \citet{Veronesi2021} fit the Elias 2-27 rotation curves obtained from the $^{13}$CO and C$^{18}$O data from \citet{Paneque2021} with two analytical models. The first was purely Keplerian, and the second included the contribution from the pressure and the gravitational potential, as in equation \ref{eq:orbital_velocity_total}. Likelihood ratios demonstrated that the data is consistent with a disk-to-star mass ratio of $0.15\lesssim q \lesssim 0.22$ which is above the threshold for GI to be active. This result, coupled with the detected GI Wiggle in \citet{Paneque2021}, is encouraging evidence in favour of GI in the Elias 2-27 system.

\begin{figure*}
  \includegraphics[width=\hsize]{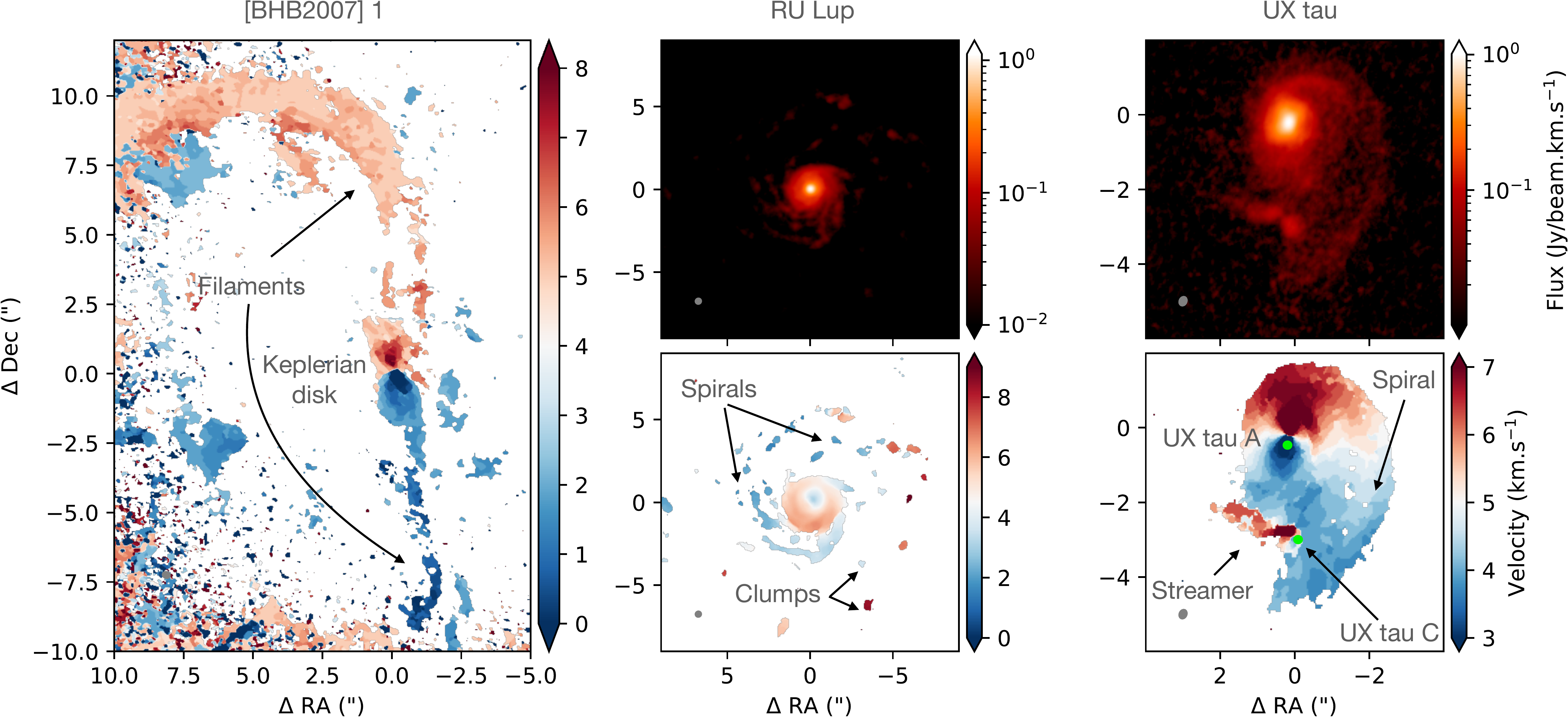}
  \caption{\label{fig:large_scale} Examples of large scale flows. Disks are not isolated systems. \emph{Left}: streamers of gas feeding a disk in [BHB2007]~1 \citep{Alves2020}. The disk display a wide gap in the continuum \citep{Alves2020}, \emph{Center}: non-Keplerian envelope, large scale spirals, and clumps in RU Lup, suggesting that GI, accretion onto the disk, or pertubations by an external companion might be at play \citep{Huang2020b}. \emph{Right:} tidal interactions from the flyby between UX Tau A and B \citep{Zapata2020,Menard2020}.}
\end{figure*}

While the Elias 2-27 system displays spiral structure in both the continuum and line emission, several systems have different morphologies when viewed in the dust continuum vs the gas. For example, the RU Lup system appears compact in the continuum, with the majority of the emission contained within $\sim$100 au and also showing prominently ringed substructure. However, $^{12}$CO observations \citep[][see also Fig.~\ref{fig:large_scale}]{Huang2020b} specifically aimed at probing the large scale structure of RU Lup revealed five blueshifted spiral arms out to 1000 au. As indicated by the brightness of the continuum emission, RU Lup is one of the most massive disks in Lupus and it is therefore possible that these spiral arms are due to GI. Two pieces of evidence support this argument: first, RU Lup is observed to have the largest accretion rate of all surveyed Class II and transitional systems in Lupus \citep{Alcala2017}, which is consistent with GI-active discs having large accretion rates \citep{Dong2015GI}. Second, RU Lup has long been known to display large photometric variations (see, eg, \citealt{gahm1974} and \citealt{giovannelli1991}). RU Lup also has 11 ``clumps'' present at distances of around $\sim$1500 au from the central star. It is possible that these clumps are consistent with fragmentation caused by GI (discussed below) and the episodic brightness variability in RU Lup is caused by the infall of these clumps \citep{vorobyovbasu2005}.

A possible outcome of GI is fragmentation, which is largely driven by cooling. Essentially, a disk that is exhibiting GI spirals is in an approximate balance between cooling and collisional heating, such that the cooling timescale $\gtrsim$ the dynamical timescale. Cooling will drive a disk that is initially stable to GI to become unstable as the Toomre parameter drops, and if it does so sufficiently rapidly the disk will fragment to form gravitationally bound objects \citep{Gammie2001,Rice2003}, with the ultimate fate of these objects largely driven by disk properties \citep{Baruteau2008,Forgan2013,Forgan2015,Hall2017,Forgan2018pop,Rowther2020}. Generally speaking, fragmentation is favoured in disks around higher mass stars \citep{Cadman2020frag} since disks tend to be more stable around lower mass stars \citep{Haworth2020}.

L1448 IRS 3B \citep{Tobin2016} is a triple protostellar system in the earliest phase of the star formation process, which increases the probability of detectable GI with an instrument such as ALMA \citep{Hall2019}. Rather than being in a self-regulating state of self-gravity, as is likely the case with the Elias 2-27 system, L1448 IRS 3B has undergone disk fragmentation into three protostellar objects objects \citep{Tobin2016}. Interestingly, recent observations using the molecular lines $^{12}$CO, SiO, H$^{13}$CO$^{+}$, H$^{13}$CN/SO$_2$ and C$^{17}$O did not show any kinematic perturbations caused by the embedded companions \citep{Reynolds2021}. This may be because the spatial and spectral resolution was below that required to detect kinematic features such as velocity kinks and the GI Wiggle, but it is also possible that the presence of the companions suppresses kinematic evidence of GI as in \citet{Rowther2020b}. If there are multiple origins of velocity perturbations within one source, careful analysis and modelling will be required to disentangle the relative velocity contributions. As demonstrated in \citet{Wolfer2021}, subtracting an azimuthally symmetric model can reveal hidden substructure in the form of large-scale spirals that is otherwise masked.

\begin{figure*}[!t]
\includegraphics[width=\textwidth]{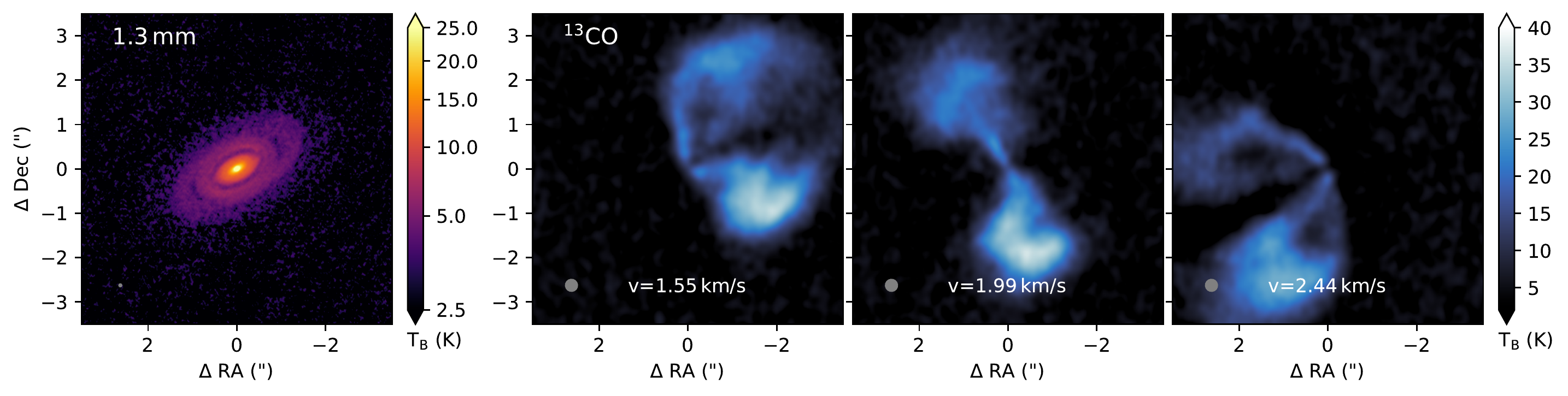}
\caption{Possible evidence of gravitational instabilities. Left panel shows the Elias 2-27 system observed in the continuum at 1.3 mm \citep{Huang2018c}. Right three panels show selected velocity channels for $^{13}$CO \citep{Paneque2021}. Strong distortions are reminiscent of the predicted GI wiggle.\label{fig:elias227}}
\end{figure*}

\section{Multiple systems and environment as traced by kinematics}

Hydrodynamic simulations have shown that star formation is a chaotic process \citep[e.g.,][]{Bate2010}. They predict a variety of protostellar disks with misaligned disks in multiple systems, successive accretion events and re-orientations, frequent dynamical encounters between protostars, erosion and tidal streams between disks, and a few disk fragmentations \citep{Bate2018,Kuznetsova2019}. Much of this is driven by the cluster environment; roughly 1 in 3 solar type stars in OB associations are expected to experience a close (100 - 1000 au) stellar flyby \citep{Pfalzner2013}, and external photoevaporation, leading to truncation of the outer disk, which can occur near massive OB stars \citep[e.g.][]{Hollenbach1994,Johnstone1998,Adams2004,Facchini2016,Winter2018}. It is interesting to notice that both of these mechanisms may have influenced the formation of the Solar System \citep[e.g.][]{Adams2010,Pfalzner2018}. In addition to environmental factors, internal perturbations can affect the disk structure; almost half of all main-sequence solar-type stars are found in binary or higher-order multiple systems \citep[e.g.][]{Duchene2013} and it is thought that the fraction of multiple systems is even higher among pre-main sequence stars \citep[e.g.][]{Duchene1999,Kraus2011,Reipurth2014_PPVI}. Stellar mass companions can have a destructive effect of protoplanetary disks \citep{Artymowicz1994,Harris2012} but multiple systems display planets and brown dwarf distributions similar to single stars \citep[e.g.][]{Bonavita2016,Asensio-Torres2018,Dupuy2016,Matson2018}, which implies that planet formation is a robust process and is indeed happening in multiple systems, including in circumbinary disks (see chapter by Offner et al.).

Observational evidence of perturbations to disks, both external (e.g. accretion along filaments, stellar flybys) and internal (binaries/planets, gravitational fragmentations) have been mounting.  \citet{deJuanOvelar2012,Mann2014, Eisner2018,Ansdell2017} found evidence that protoplanetary disks in high stellar density environments have smaller, less-massive disks (with some exceptions, see chapter by Manara et al. in this book). The importance of environmental factors appears to be most prominent for cluster densities $\gtrsim$10$^{4}$ pc$^{-3}$ \citep{deJuanOvelar2012,Rosotti2014,Winter2018}, roughly corresponding to the central stellar density in the ONC \citep{Hillenbrand1998} and other nearby star-forming regions \citep{Kuhn2015,Pokhrel2020}. Individual sources show evidence of tidal interactions in nearby star forming regions, suggesting that tidal interactions are important even in low mass star forming regions \citep{Pfalzner2021}. Often a single observed feature (e.g., a thin stream of gas stretching away from the disk) can have multiple potential explanations (e.g., a jet being launched by the star, late-stage accretion onto the disk, flyby from a stellar/cloudlet interloper), but kinematics provide a valuable diagnostic for distinguishing these origins. We will review these observations and discuss how disk dynamics, and potential for planet formation, is affected, and how in turn disk kinematics can be used to trace these processes.

\subsection{Streamers, stellar fly-bys and tidal interactions}
\label{sec:streamers}

Observational imprints of the chaotic star formation process can be seen in the structures surrounding young protostars (see chapter by Pineda et al. for a detailed discussion). Tails of molecular line emission, stretching $\sim100-10000$ au, have been detected around the protostars HL Tau \citep{Yen2019}, Per-emb-2 \citep{Pineda2020}, [BHB2007]~1 \citep{Alves2020}, and IRAS 16293-2422 A \citep{Murillo2022}, thus showing that infall of material along narrow filaments on early disks does occur. Accretion shocks from infalling material have also been observed using SO and SO$_2$ emission around DG Tau and HL Tau \citep{Garufi2022}. Interestingly, the class II SU Aur system exhibits a $\sim$1000 au tail that is seen in scattered light \citep{deLeon2015,Akiyama2019} and CO \citep{Ginski2021}. The molecular line observations find velocities of $\sim1-2$ km s$^{-1}$ for the tails, with little to no velocity gradient, indicating that they are not jet features being launched by the central star.

The combination of the gas kinematic data and the scattered light observations in the IR allowed \citet{Ginski2021} to conclude that the large scale tail is streaming towards the disk, rather than being ejected by a ballistic encounter with an undetected companion. This indicates that late stage infall can occur also in Class II sources \citep[e.g.][]{Kuffmeier2017}. The case of SU Aur could be similar to AB Aur, where the large spirals detected in scattered light within the disk \citep{Fukagawa2004,Boccaletti2020} may indicate the inflow of low angular momentum material accreted via infall from the surrounding low density gas \citep[e.g.,][]{Hennebelle2017,Kuffmeier2020}. The possibility of successive accretions of molecular ``cloudlets'' has also recently been explored \citep{Dullemond2019,Kuffmeier2020}. They may also lead to the formation of misaligned second-generation disks that can survive for more than 100 kyr \citep[][see also section~\ref{sec:warps}]{Kuffmeier2021}.  Other systems show very faint large scale spirals in the outer disk that are reminiscent of low density material being slowly accreted at the outer edge of these objects \citep[e.g.,][]{Grady1999,Fukagawa2004,Ardila2007,Huang2020b,MAPS19}.

In the case of Per-emb-2 the gas tail extends to $\approx$ 10\,000\,au and is massive enough (0.1-1M$_{\odot}$, compared to $\sim$3M$_{\odot}$ for the core itself) to substantially influence the mass and angular momentum of the disk. Filaments can increase the accretion timescale for protostellar cores, given that the free-fall time increases with radius, and with sufficient mass could change the angular momentum of the disk itself \citep{Kuffmeier2017}. It is intriguing that SU Aur exhibits evidence of a misaligned inner disk through a sharp double-sided shadow (see Section \ref{sec:warps}), suggesting that the outer disk is being misaligned by accreting material with a different angular momentum direction from the infall \citep{Ginski2021}, similarly to what is observed in [BHB2007]~1 \citep{Alves2020}.

Infall is not the only possible cause of extended tails, as in the case of RW Aur where the observations of its $\sim$600 au one-armed spiral \citep{Cabrit2006,Rodriguez2018} and the 50 years relative astrometry of the binary system can be reproduced by a fly-by between the two components of the system \citep{Dai2015}. A large fraction of the disk around the primary star is ejected along a spiral trailing the disk. A study of the gas kinematics shows that the projected velocities of the spiral are well above the escape velocity limit \citep{Cabrit2006}, thus showing the disruptive nature of close encounters, in line with hydrodynamic simulations \citep{Clarke1993,Munoz2015,Dai2015,Cuello2019,Cuello2020}.

Additional multiple systems show signs of stellar interactions, such as tidal streams: FU Ori \citep{Takami2018,Perez2020}, UX Tau \citep{Menard2020,Zapata2020} and Z~CMa \citep{Liu2016,Dong2022}; disk-disk interactions: AS 205 \citep{Kurtovic2018}, HV \& DO Tau
  \citep{Winter2018}; externally excited spirals: HD 100453
  \citep{vanderPlas2019,Rosotti2020b}. Several multiple systems show evidence for misalignment between different components \citep[e.g.][]{Jensen2014,Tobin2018,Czekala2019,Manara2019}. Tidal structures are also detected at earlier phases (class I, \citealp{Tobin2016,Alves2019}). The observed features are consistent with numerical predictions \citep{Dai2015b,Bate2018,Cuello2019,Vorobyov2020} and offer laboratories to better understand the disk dynamics by quantifying their response to gravitational torques. In particular, the synergy of gas dynamics in spirals, and the distribution of small and large dust grains as traced by IR scattered light images and mm thermal emission, can lead to constraints on the coupling between gas and dust in disks, and on the thermal structure of the disk itself \citep{Juhasz2018,Rosotti2020b}. All these interactions can affect the accretion on the central star \citep[e.g.][]{Tofflemire2017}.

It is difficult to determine the frequency of late-infalls and star-disk encounters in class II disks, in particular due to the lack of uniform and deep large scale observations of molecular gas at typical scales of $\sim$1000 au. Nonetheless, serendipitous detections show that both tidal interactions and late infalls seem to be much more common than previously considered in low density star forming regions \citep{Winter2018,Pfalzner2021}. This may be due to higher densities in low-mass clusters than previously used in $N$-body simulations of low stellar clusters \citep{Pfalzner2021}, or to a  chaotic star formation phase with relic gas affecting disk evolution for long timescales. Whatever the cause, their effects could play a crucial role in determining the planet formation outcome of several systems, affecting the mass and angular momentum evolution of disks during their planet formation phase.

\subsection{FU Ori mechanism}
\label{sec:FU_Ori}

FUor outbursts are characterized by a sudden increase in brightness, often at optical and infrared wavelengths, of over an order of magnitude followed by a slow decay over decades (see \citealp{Audard2014} for a review). These bursts in luminosity are attributed to increases in the accretion rate, although the exact mechanism by which the accretion rate increases by orders of magnitude over the course of a few months is unknown. Theories include internal mechanisms, such as the combination of magneto-rotational instabilities and gravitational instabilities \citep[e.g.][]{Armitage2001,Zhu2009,Bae2014}, the fragmentation and infall of gravitational instabilities \citep{Vorobyov2010,Vorobyov2015,Machida2011,Meyer2017}, or planet-driven thermal instabilities \citep[e.g.][]{Lodato2004}, as well as external mechanisms such as binary interactions \citep{Reipurth2004}, and star-disk flybys \citep{Pfalzner2008a,Pfalzner2008b,Cuello2019,Borchert2022,Dong2022}. The wide variety of observed FUOr outburst light curves makes it difficult to isolate a single driving mechanism.

\cite{Vorobyov2021} explore the possibility of using disk kinematics to constrain the burst mechanism. They find that MRI+GI drives deviations in the orbital and radial motion that are a few percent of the Keplerian velocity, while the infall of a gravitational clump creates disturbances of tens of percent of the Keplerian velocity. A stellar intruder not only drives more powerful outbursts, it also creates deviations in orbital and radial motion that are comparable to, if not larger than, the Keplerian velocity.
\cite{Borchert2022} showed that a stellar flyby can indeed trigger a long lasting ($\approx$ 100 yr) FU Ori like outburst with a fast rise time ($\approx$ 1yr) as long as the secondary penetrates the circum-primary disk. In that case, the outburst occurs on the secondary, and the short rise time is a consequence of circum-primary disk material falling directly on the secondary star.

ALMA observations of FUor objects have mainly focused on the continuum, finding massive compact disks \citep{Zurlo2017,Cieza2018,Kospal2020}, although observations have begun to explore the kinematics. Disk-like kinematics have been detected around V346 Nor \citep{Kospal2017}, V883 Ori \citep{Ruiz-Rodriguez2017}, and both components of the FU Ori binary \citep{Perez2020}, although not at the spatial and spectral resolution needed to discern deviations from Keplerian motion. FU Ori does show arc-like and extended features in CO that may be signs of binary interaction, albeit among complicated outflow/envelope emission and foreground cloud absorption \citep{Hales2015}. V2775 Ori displays blue and re-shifted rings, that have been interpreted as conical flows \citep{Zurlo2017}, but may also be the signatures of dynamical interactions with a flyby \citep{Cuello2020,Borchert2022}. \cite{Dong2022} showed that a flyby can explain the 2000\,au streamer detected in scattered light and CO emission in the Z~CMa system.

\subsection{Misaligned inner disks, radial flows and kinematics of transition disks}

\label{sec:warps}

Disk warps, or variable disk orientation with stellocentric radius, have often been proposed in the context of young circumstellar disks, generally in the form of tilted inner disks that are misaligned relative to the outer regions. Numerical simulations have suggested that a misaligned companion or planet could warp or even break the disk in two or multiple precessing rings \citep{Nixon2013,Facchini2013,Teyssandier2013,Lodato2013,Nealon2018,Facchini2018,Zhu2019}.
Infall events occurring after the initial collapse can also result in misaligned outer disks \citep{Kuffmeier2021}.
Late type stars are also known to possess a strong magnetic dipole moment, due to their convective nature. When the magnetic field axis is misaligned with respect to the disk inclination, the ionized inner disk can also become warped and slowly precess \citep[e.g.][]{Foucart2011,Romanova2021}.
Radiative transfer models of the hydrodynamical simulations showed that disk warping, or breaking, can be probed by the shadowing cast by the inner disk onto the outer regions of the disk itself, as traced by scattered light or thermal emission \citep[e.g.,][]{Marino2015,Min2017,Juhasz2017b,Facchini2018,Nealon2020b}.

 Post processing of the gas component of the hydrodynamical simulations shows that disk warping or breaking can also be traced by disk kinematics, which can be used to trace a gradient in the inclination and position angle of a rotating disk (see Section~\ref{sec:methods:rotation_curves}).

The case of misaligned planets (or low mass stellar companions) is particularly interesting. In particular, a detailed comparison of observations and hydrodynamical simulations can help inferring the inclination angle and radial location of a purported perturbing body in disks that would be otherwise unnoticed \citep[e.g.,][]{Lodato2013,Price2018,Nealon2019,Nealon2020c,Ragusa2021,Ballabio2021}. Among others, the interesting prediction of an inner disk that is tilted by the secular perturbation of a misaligned companion is that the inner disk would be precessing. If the precession timescale is short enough, the movements of the shadows cast by this inner disk onto the outer regions can be observed within a human lifetime, with a peculiar pattern that depends on the geometry of the system \citep{Facchini2018,Pinilla2018,Nealon2020b}. Such evolution can also be used to infer the mass and orbital radius of an unseen companion.

Strong evidence
  for such warps was provided thanks to advances in high-contrast and
  polarimetric imaging in the near-IR, which revealed that the
  detailed shape of the intensity decrements along the ring of
  HD\,142527 \citep[][]{Fukagawa2006,Casassus2012,Avenhaus2014}
  matched that of the two-sided shadow cast by a inner disk tilted at
  a dramatic angle of $\sim$70\,deg \citep[][]{Marino2015}.  The case
  of disks with large central cavities,
  such as HD\,142527, is particularly useful to pick up and
  characterize pronounced inner disk tilts since their rings are
  directly exposed to stellar light and are very bright in reflected
  light, except for the regions that are shadowed by the inner disk.
  The shape of such shadows results from the projection
  of the flared inner disk surface onto that of the outer edge of the
  cavity, and extends into the outer disk. Geometrical arguments can
  connect the position of the shadows along the rings with the
  orientation of the inner disk \citep[][]{Min2017}. Further examples
  of such sharp, narrow and two-sided shadows have been observed in
  the disks of HD\,100453 \citep[][]{Casassus2016, Benisty2017},
  J1604-2130 \citep{Mayama2018, Pinilla2018},
  DoAr\,44 \citep[][]{Casassus2018}, HD~135344~B \citep{Stolker2016,Stolker2017}
and GG~Tau \citep{Keppler2020}.

  Shallower inner disk tilts have also been invoked to explain large
  scale scattered-light modulations in azimuth, as for instance in the
  low amplitude modulation in HST images of TW\,Hya
  \citep[][]{Debes2017}, or in the SPHERE polarimetric
  images of HD\,143006 \citep[][]{Benisty2018}. The
  case of the wide shadow in HD\,139614 is particularly interesting,
  as it requires at least three disk planes or large gradients in the position angle of an inner warp
  \citep[][]{Muro-Arena2020a}.

  Inner disk tilts have also be inferred from continuum images of
  thermal radiation. This is usually performed by comparing the outer
  disk orientation with that of optical-IR interferometric
  observations of the inner disks. For instance,
  \cite{Bouvier2020} confirm the inner disk
  orientation inferred from the scattered-light shadows in
  DoAr\,44. High-resolution sub-mm continuum observations with ALMA
  have beautifully exposed a large-scale warp in GW\,Ori
  \citep{Kraus2020,Bi2020}, whose triple ring system is seen as a
  projected ellipses with different inclinations. \cite{Perraut2021}
 detected clear misalignments between both disks for 4 objects out of 9 transition disks, while \cite{Bohn2022} detected additional misalignments in 6 targets out of sample of 20 transition disks. For 3 of those targets the orientation of the inner disk is consistent with the shadows detected in scattered light in the outer disks. Slight misalignments are not ruled out for the remaining objects.

  An intriguing feature of protoplanetary disk warps is the apparent
  dichotomy in the thermal response of the shadowed regions. Some
  optical-IR shadows correspond to (shallow) drops in sub-mm continuum
  emission, such as HD\,142527
  \citep[][]{Casassus2015b}, J1604
  \citep[][]{Mayama2018} or DoAr\,44
  \citep[][]{Casassus2018}, while most do not. The depth of the
  continuum decrements traces a corresponding drop in the dust
  temperature, which may be too shallow to detect if the shadow
  crossing time is too short compared to the cooling timescale of the
  shadowed regions. In addition, the thermal radiation from the disk
  itself can also heat the shadowed regions
  and smooth out any temperature decrements, especially for optically thin regions (such as in lower mass outer disks). Simple models for the thermal response of the shadowed regions are proposed in \citet{Casassus2019}.

  At the time of writing the frequency of warps in protoplanetary
  disks is not yet fully understood. However, the light-curve
  variability of the so-called ``dipper stars'', i.e. the AA\,Tau-like
  classical T-Tauri stars, offers some insight, since such stars are
  thought to be viewed edge-on.  As shown by \citet[][]{Ansdell2020},
  the statistics of the outer disks orientation in dipper stars,
  inferred from ALMA continuum observations, appear to be uncorrelated
  with that of the inner regions. Protoplanetary disk warps would
  therefore be quite common, with a probably uniform
  distribution of inner disk tilt angles.

  When available, the molecular line kinematics traced in the
  rotational lines of $^{12}$CO are consistent with the warp
  geometries inferred from scattered-light imaging, as long as the
  tilted inner regions can be resolved, such as is possible with ALMA
  observations. The case-study of HD\,142527 is an extreme, with fast
  accretion through the sharp warp \citep{Casassus2015a}
  reaching free-fall values before braking near the inner disk, and
  then accelerating away from the outer disk plane and into the plane
  of the inner disk, with vertical velocities that are also comparable to
  Keplerian. The fast intra-cavity flows in HD\,142527 are
  probably driven by
  HD\,142527B \citep{Biller2012}, as demonstrated  hydrodynamical simulations of an inclined inner binary explain most of the observed features of the disk \citep{Price2018}.  Warped kinematics are also observed
  near the inner disk in J1604 \citep{Mayama2018} and in
  HD\,143006 \citep[][]{Perez2018b}, GG Tau \citep{Phuong2020}. Gas kinematics can also reveal so far unseen companions in the cavity of transition disks \citep{Calcino2019,Poblete2020,Calcino2020}. While current estimates suggest that binaries can explain $\approx$ 40\% of the transition disks \citep{Ruiz-Rodriguez2016}, kinematic studies at high spatial resolution might be used to refine these estimates and better understand the flows of gas between outer and inner disks, as well as the high accretion observed in the brightest transition disks.

One of the theories for the formation of the azimuthal dust concentrations seen in sub-millimeter continuum emission invokes anticyclonic vortices \citep{Barge1995,Lyra2009,Meheut2010,Meheut2012,Lyra2013,Baruteau2016,Sierra2017}. They can form at low viscosity ($\alpha \lesssim 10^{-4}$) via the Rossby wave instability \citep[e.g.][]{Lovelace1999,Li2000} when steep gradients in density \citep{deVal-Borro2007,Zhu2014} or in viscosity \citep{Varniere2006,Regaly2012} are present. The edge of a disk cavity of transition disks is a favorable site for vortex formation.
In the context of transition disks with large cavities, \cite{Huang_p2018}, \cite{Perez2018} and \cite{Robert2020} showed that the counter-rotation of a vortex with respect to the disk is faint but detectable with a careful choice of spatial and spectral resolutions. A detailed analysis by \cite{Boehler2021} showed that the non-Keplerian motions detected near the dust concentration in HD~142527 might be consistent with the kinematic signatures of a vortex. However, at the current resolution, they remain indistinguishable from artificial velocity deviations generated by beam smearing.

\section{The kinematics of planet-disk interactions}
\label{sec:planet-disc_interactions}

The search for young, forming, planets has been focused on detecting emission associated with the planet itself, either via their direct emission, thermal emission associated, or signatures of accretion. The spectacular case of the PDS~70 system should not occult how challenging such detections have proven to be. Young planets are embedded inside their parent circumstellar disk, and as discussed earlier, in most cases this leads to significant optical depth towards the observer, basically hiding the planets. The ubiquitous dust sub-structures seem in disks, at both near-infrared and sub-millimetre wavelengths, suggests that young planets are numerous in disks by the time we observe them, but multiple mechanisms can explain the observed sub-structures (see chapters by Bae and Paardekooper).

With the capability of ALMA to image molecular emission at both high spatial and high spectral resolution, an alternative and complementary approach has emerged, searching for kinematic signatures of embedded planet \citep{Disk-Dynamics-Collaboration2020}. In this section, we will review the theoretical expected kinematic signatures of embedded planets, and present the current state of the art in terms of detecting planets through their gravitational influence on the dynamical structure of their parental disk. With multiple tracers, we are starting to be able to trace the dynamical structure of the gas at a range of heights in the disk, offering new ways to distinguish between potential mechanisms which are driving the substructures in the dust. We provide a set of criteria which can be used to claim a detection of an embedded planet.

\begin{figure*}
  \centering
  \includegraphics[width=\hsize]{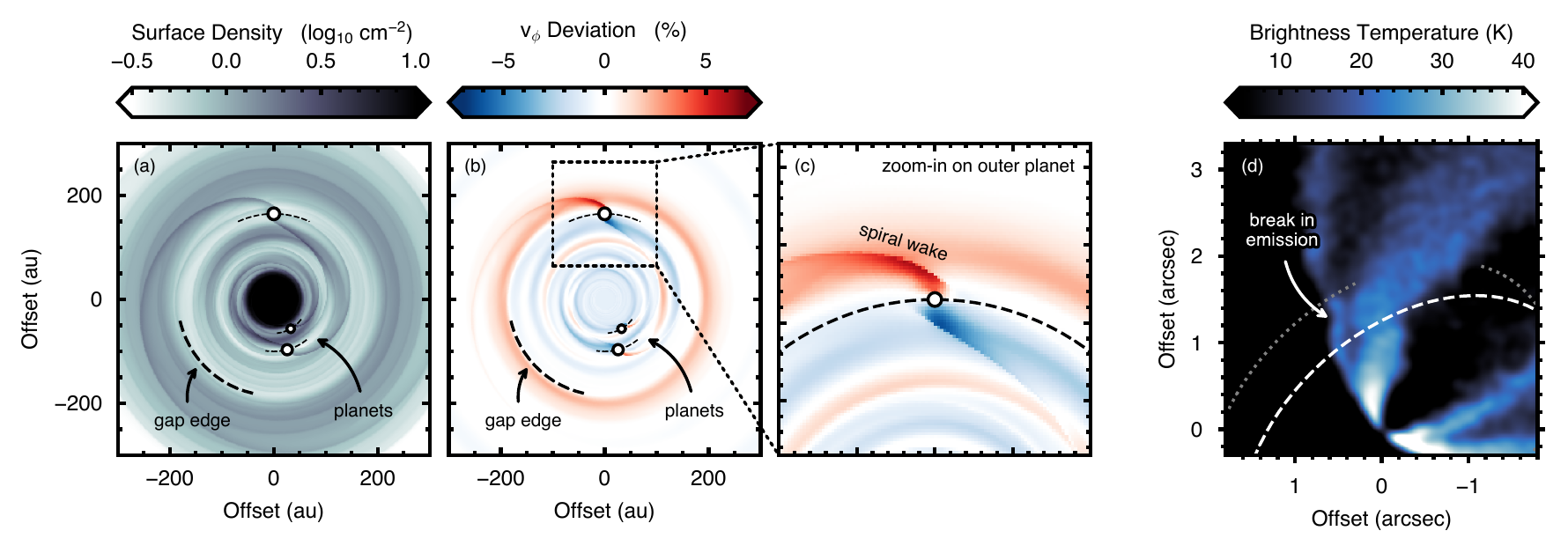}
  \caption{
    Numerical simulations of planet-disk interactions for a multi-planet system based on HD~163296 hosting 3 Jupiter-mass planets. (a) shows the gas surface density; planets open gaps in addition to exciting spiral arms. (b) and (c) show the deviation in rotation velocity caused by the planets, with a zoom-in to the outer planet. The most notable features tracing the gap edges and spiral wakes. (d) shows the simulated emission morphology with a characteristic `kink' induced by the velocity perturbation caused by the orbiting planet. Here the disk and planet orbit are oriented as observed for HD~163296. The dotted line shows the orbit along the midplane, while the dashed line shows the orbit along the CO emission surface.}
  \label{fig:planetdiskinteractions}
\end{figure*}

\subsection{Expected kinematic signatures of embedded\\ planets}
\label{sec:planet-disc_interactions:expectations}

\subsubsection{Spiral wakes}
\label{sec:spirals}
Embedded planets exert a gravitational perturbation on the surrounding disk, which can lead to an exchange of energy and angular momentum. Planets  launch spiral density waves at Lindblad resonances both inside and outside their orbits \citep{Goldreich1979,Goldreich1980,Papaloizou1984}. The superposition of the waves form a one-arm spiral density wave, called the planet wake \citep{Ogilvie2002}. The wave propagates away from the planet, eventually becoming non-linear and steepening into shocks \citep{Goodman2001,Rafikov2002}. The planetary wake disturb the local density structure of the disk, but also velocity field. \cite{Goodman2001} and \cite{Rafikov2002} presented a semi-analytical description of the structure of the density perturbation for a disk with an adiabatic equation of state. This work was extended by \cite{Bollati2021} to describe the associated velocity field. The gravitational pull of the planet will induce deviation from Keplerian rotation of up to a few 10\,\%, with the inner wake being sub-Keplerian and the outer wake being super-Keplerian. In addition to changes to the azimuthal velocities, the spiral wakes also result in radial motions of comparable amplitudes \citep[e.g.][]{Pinte2019}. Deviations are mainly azimuthal near the planet, while radial motions dominate far away along the the wake \citep{Rafikov2002}.  Vertical motions in the form of meridional eddies are also associated with the wake \citep{Fung2016}. The exact shape and amplitude of the velocity deviation depends on the planet mass, and disk structure, in particular its thermal structure, \emph{i.e.} its aspect ratio. Both semi-analytical  \citep{Bollati2021} and numerical \citep{Rabago2021a} calculations indicate  that the velocity deviations depend linearly on the planet mass in the vicinity the planet, and they scale with the square root of the planet mass further away. Both also show that there is little dependence of the amplitude of the velocity deviations on the disk viscosity.

The disturbed velocity pattern is detectable with high spectral and high spatial resolution ALMA molecular line observations.
This possibility was first envisaged by theoretical work that predicted the observable features associated with embedded planets \citep[e.g.][]{Perez2015,Perez2018}.
In a given channel map, the emission is concentrated along the iso-velocity curve, \emph{i.e.}, the region of the disk where the projected velocity is constant. In the presence of a planet, the gas flow is perturbed and an additional Doppler shift in the molecular emission is induced. As a result, the iso-velocity region is distorted and the emission displays a distinctive ``kink''.

On top of Lindblad's spirals, planets can also excite buoyancy spirals if some thermal conditions are met in the disk \citep{Zhu2012,Zhu2015,Lubow2014,McNally2020}. Because of the vertical temperature gradient in disks, a spatial perturbation of a gas parcel leads to a vertical oscillation around its equilibrium position if the adiabatic index is larger than 1. For this perturbation to develop into a buoyancy resonance, the timescale for the gas to respond to thermal perturbations needs be longer than the timescale associated with the buoyancy. \cite{Bae2021} argue that the slow thermal relaxation in the disk surface due to infrequent gas/dust collision provides favorable conditions for buoyancy resonances to develop. Interestingly, buoyancy spirals are tightly wound compared with Lindblad spirals, in particular in the vicinity of the planet. They also produce predominantly vertical motions, leading to different kinematics signatures from Lindblad's resonance spirals.

\cite{Muley2021} showed that the shock generated by the spiral wake can additionally led to a temperature increase in the spiral if the cooling timescale exceeds the dynamical timescale. When the cooling is slow, the  temperature perturbations from buoyancy spirals can be similar to those from Lindblad spirals. Such an increase in gas temperature can be expected to be detected both in term of line flux and line width.

\subsubsection{Gaps}
The spiral wakes generated by the planet will exert a torque on the disk (positive for the inner wake and negative for the outer wake). If this torque is larger than the viscous torque, the planet will carve an annular gap along its orbit \citep{Lin1986}.
The steep pressure gradients at the edge of the gaps will modulate the azimuthal velocities, leading to a super-Keplerian flow at the gap outer edge and a sub-Keplerian flow at the inner edge (\citealp{Kanagawa2015,Perez2015,Gyeol_Yun2019}, see also Fig.~\ref{fig:azimuthal_velocities}).
 Note than small mass planets can open a gap in the dust structure while only mildly impacting the gas profile \citep[][see chapter by Lesur]{Dipierro2016}, and the presence of a gap in the sub-millimeter continuum emission does not imply that they will be associated with deviations from Keplerian rotation.

The reduced opacity associated with gaps (in particular dust gaps) can also lead to sharp variations of temperature as the stellar radiation can penetrate the disk deeper inside the gap. This can impact the pressure gradient, and in the case of deep dust gap, the temperature gradient can dominate, potentially hiding any gas density gradient or gap \citep{Rab2020}.

 \subsubsection{Meridional flows}
 The gap profile is an equilibrium between the planet torque and the viscous and pressure torques.
 Because of the vertical extent of the gas disk near the planet, the asymmetry in the torque drives significant motions, in particular ``meridional flows'' \citep{Kley2001,Tanigawa2012,Szulagyi2014,Morbidelli2014}. As the planet opens the gap, gas viscously spreads towards the gap center and towards the midplane, creating eddies in the gas at the edges of gap. It remains unclear how extended those flows are around the planet, but most simulations indicate that they are the strongest near the planet. Note that these meridional flows should appear for any gap in the gas surface density, independently of the physical process creating the gap, and the azimuthal variation of such flows might provide insight on the underlying physical process causing the gap.

\subsubsection{Turbulent motions in the gap}
On top of those ordered flows, an embedded planet can drive transonic to mildly supersonic vertical motions \citep{Dong2019}, which should result in a local enhancement of the line width. Interestingly, these turbulent motions appear at all azimuth in the gap, and can potentially help to distinguish between a planet and other mechanisms as the origin of the gap. They are likely challenging to detect however as the gap needs to be spatially resolved, and the observations deep enough to extract a robust line width, as well as to determine accurately the gas temperature in order to measure non-thermal motions.

\subsubsection{Circumplanetary disks}

At much smaller scales -- about 1/3 of the Hill radius --  the accretion of material onto the planet is regulated by the formation of a circumplanetary disk \citep[CPD,][]{Kley1999,Lubow1999}. The exact nature of the accretion process onto the planet remains poorly known, and models predicts a range of size and temperature for the CPD depending on the local equation of state and adopted opacities \citep{Bate2003,Dangelo2003,Szulagyi2014,Szulagyi2017}. These simulations suggest that a CPD will form for planet more massive than a Saturn mass. Even with a 10 M$_\mathrm{Jup}$ planet, detecting the CPD requires deep observations at high resolution \citep{Szulagyi2018,Zhu2018}. The recent detections of the CPDs around PDS~70~b in the near-infrared \citep{Christiaens2019} and  around PDS~70~c in the sub-millimetre \citep{Isella2019,Benisty2021} have confirmed their existence.

 The nature of gas flow in the vicinity of the CPD also varies depending on the underlying assumptions made in the model. But, in any case this accretion onto the planet will add a large velocity component to the gas. \cite{Perez2015} showed that the velocities of these flows are large enough for the CPD emission to decouple from the background disk emission, resulting in a spot detached from the expected iso-velocity curve of the circumstellar disk. Because the CPD itself is not resolved, its emission is also associated with a local increase in velocity dispersion.
 Recently, \cite{Martin2020} investigated the conditions under which a CPD can detach from the disk and its tilt can grow, arguing that a rapid growth is typically favoured. This may favour detectability of the CPDs depending on the respective orientations of the circumstellar and circumplanetary disks relative to the line of sight.

\subsubsection{Anticyclonic vortices}
\label{sec:vortex}
Similar to the edge of a cavity in a transition disk,  the edge of a ring carved by a planet is a favorable site for the development of vortices \citep[e.g.][]{Baruteau2019}.  It remains unclear whether a smaller vortex generated by a planet might be detectable in kinematics, in particular as the same artefact due to beam smearing as in the case of transition disk will apply \citep{Boehler2021}.

\subsection{Kinematic detections of embedded planets}

\begin{figure*}
  \includegraphics[width=\hsize]{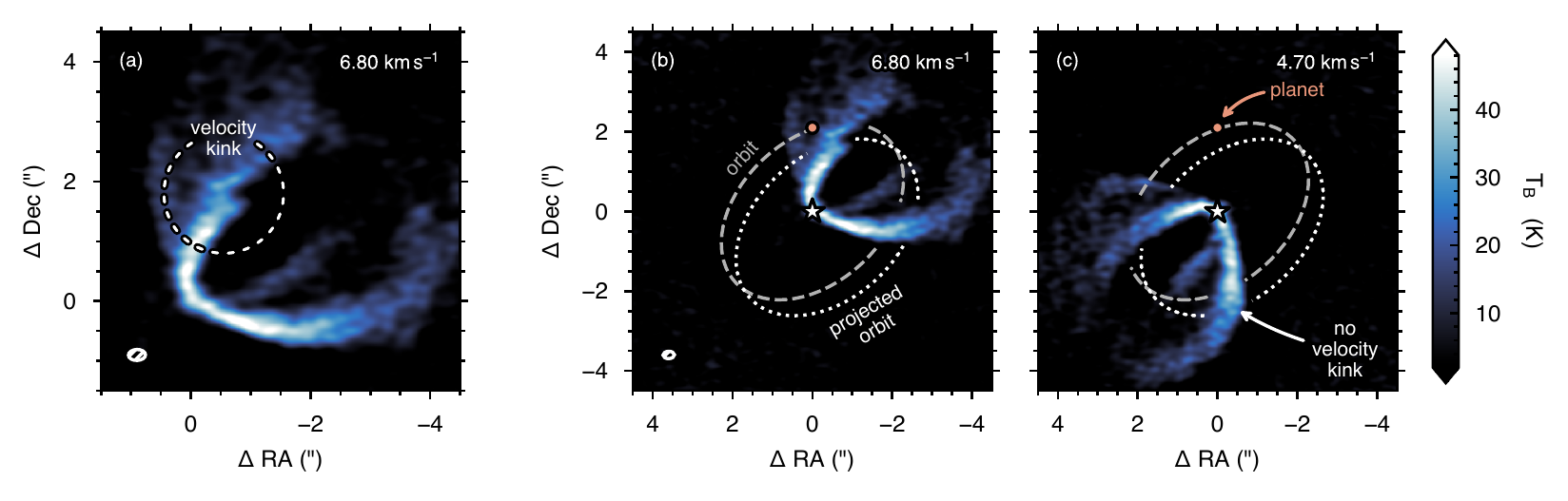}
  \caption{Kinematic asymmetry in the $^{12}$CO J=2-1 line emission of HD~163296 (left, \citealp{Pinte2018b}). Comparison with the channel on the opposite side of the disc (right) shows no corresponding feature, indicating
    the disturbance to the flow is localized in both radius and
    azimuth. The inferred planet location is marked, as well as its orbit and its projection at the height traced by $^{12}$CO.
  \label{fig:HD163}}
\end{figure*}

\begin{figure*}[!t]
  \includegraphics[width=\textwidth]{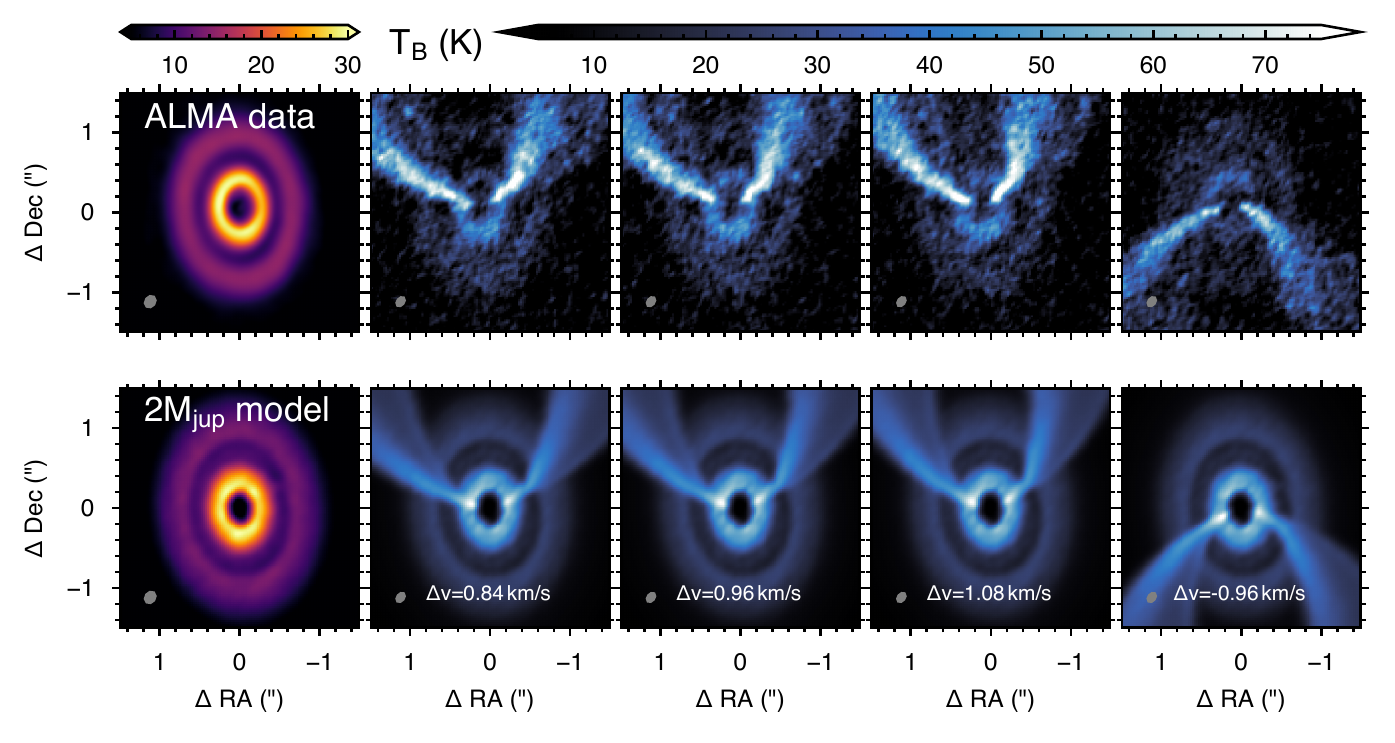}
  \caption{Kinematic detection of an embedded protoplanet carving a dust gap. Top: ALMA observations of dust (left panel) and gas kinematics (right panels) in the HD~97048 protoplanetary disk. The kink evident in the velocity channel maps is evidence for an embedded planet that also carves the gap seen in the dust. Bottom: 3D hydrodynamics model of a newborn planet embedded in a protoplanetary disk, computed with phantom and post-processed with the mcfost radiative transfer code. By matching models to observations we are able to directly constrain the planet mass (here best fit by a planet twice the mass of Jupiter). Adapted from \cite{Pinte2019}.\label{fig:kink_model}}
\end{figure*}

Excitingly, the possibility of kinematically detecting planets has become reality. Detections of embedded planets have already been claimed from distortion of the iso-velocity curves in individual channel maps in HD~163296 \citep[][see Fig.~\ref{fig:HD163}]{Pinte2018b}, and in HD~97048 \citep{Pinte2019}. In both cases, the kinks are localized in velocity and space, excluding any large structure in the the disk, and suggesting a local origin. The excellent agreement with predictions from simple hydrodynamical models of a giant planet in a disk (Fig.~\ref{fig:kink_model}) led to the rapid adoption of embedded planets as the explanation for the kinks observed in high resolution ALMA images. Synthetic channel maps show that the kinematic signature strongly depends on the planet mass \citep[e.g.][]{Perez2015}. By comparing a range a models exploring the planet mass, while keeping the disk parameter fixed, it is possible to obtain constraints on the mass of the embedded planets. For HD~163296 and HD~97048, \cite{Pinte2018b} and \cite{Pinte2019} showed that in both cases the planets are around 2 to 3 Jupiter masses.

For HD~163296, the planet is located at 260\,au from the central star, inside a gap detected with HST/STIS \citep{Grady2000,Rich2020}, which also corresponds to a dip in the integrated $^{12}$CO emission \citep[][also visible in the top right panel of Fig.~\ref{fig:channel_maps}]{Isella2016}. \cite{MAPS18}, \cite{Izquierdo2022} and \cite{Calcino2022} showed that the velocity kink is associated with a spiral structure that spans most of the disk. In particular,  \cite{Calcino2022} demonstrated that the spiral structure is the outer planetary wake generated by the planet.
\cite{MAPS18} and \cite{Izquierdo2022} detected kinematic structures at $\approx$ 145\,au suggesting a potential additional planet.

In the case of HD~97048, the planet is located inside the dust gap detected in sub-millimeter continuum emission at 130\,au, demonstrating that embedded protoplanets are responsible for at least some of the gaps observed in disks. Similarly, tentative detections were presented by \cite{Pinte2020} in 8 out of 18 DSHARP sources \citep{Andrews2018}, as well as a tentative detection in CI~Tau by \cite{Rosotti2021}.  In most cases, the deviations from Keplerian rotation are localized and coincide with a gap, pointing towards potential embedded planets, but the limited signal-to-noise of the data and foreground $^{12}$CO prevents a definite answer in those objects.

By subtracting the Keplerian rotation, the additional velocity component from the spiral wake created by the presence of a planet can manifest as a ``Doppler-flip'' \citep{Perez2018}, \emph{i.e.} a sign-reversal in the non-Keplerian component of the gas velocities. The Doppler-flip in the rotation maps and kink in individual channel maps are two equivalent signatures of the same underlying velocity structure. Depending on the signal-to-noise, spectral and spatial resolution, as well as the geometry of the system (orientation of the disk and azimuth of the planet), the signature of an embedded planet might be easier to detect in one or the other representation of the data cube \citep{Pinte2020}. \cite{Casassus2019b} and \cite{Perez2020b} reported the detection of such a feature in the disk of HD~100546, a ``transition'' disk with a bright dust ring at $\approx$ 20\,au.  The amplitude of the velocity deviations in HD~100546 (detected over 7\,km/s and 90$^\circ$ in azimuth) suggests the responsible body might be more massive than a planet.
The Doppler-flip coincides with complex structures in the  ring, and seems to align with intricate grooves in the  dust continuum emission,  hinting at a complex dynamical scenario. Non-axially-symmetric velocity structures are also present inside the cavity, but at lower amplitude than the Doppler-flip. The faintness $^{12}$CO emission prevents testing for the presence of a tilted inner disk as in HD~142527.

\cite{MAPS18} revealed a highly structured disk around MWC~480 with concentric radial variations in temperature, as well as low pitch angle spiral perturbations. These structures appear broadly consistent with the predictions from buoyancy spirals, and may indicate the presence of planet located at $\approx$ 245\,au.

For more face-on sources, the projection of the velocities along the line-of-sight velocities result in different components of the velocity perturbations being traced, with a strong weighting to the vertical motions. Using $^{12}$CO (3-2) observations of TW Hya, \citet{Teague2019b} revealed a large, localized velocity deviation with projected deviations of ${\approx}~40~{\rm m\,s^{-1}}$ associated with tightly wound spirals traced by variations peak intensity. The coincidence of this deviation with the large, gas- and dust-depleted gap at ${\approx}~90$~au \citep{vanBoekel17, Teague2017} strongly suggested that the cause of the velocity perturbations was likely responsible for opening the gap. \citet{Bae2021} demonstrated that tightly wound and highly localized velocity deviations could be buoyancy resonances and that a small, $< 0.5~M_{\rm Jup}$ planet could be responsible for such deviations. Follow-up observations at a higher spectral resolution will be necessary to confirm these findings.

\subsection{Mapping the 3D gas flow in the vicinity the planet}

As discussed in Section~\ref{sec:planet-disc_interactions:expectations}, planets are expected to drive a range of kinematic features both in their immediate vicinity and across large regions of the disk. In order to differentiate these features the projected line-of-sight velocity must be decomposed into its disk-frame cylindrical components: $v_{\phi}$, $v_r$ and $v_z$. The projection of these components along the line-of-sight is given by
\begin{align}
    v_{\phi,\, {\rm proj}}  &= v_{\phi} \cos(\phi) \sin(|i|),\\
    v_{r,\, {\rm proj}}     &= v_{r} \sin(\phi) \sin(i),\\
    v_{z,\, {\rm proj}}     &= -v_{z} \cos(i),
\end{align}
where $\phi$ is the polar angle in the disk frame (such that $\phi = 0\degr$ corresponds to the red-shifted major axis) and $i$ is the inclination of the disk. In this formalism we allow $i \in (-90\degr,\, +90\degr)$ to account for the fact that the disk is three dimensional, but we only trace one surface. Here positive $i$ represent a disk that is rotating in a counter-clockwise direction, while negative $i$ described a clockwise rotating disk. Therefore, if the geometry of the disk can be constrained then any velocities can be deprojected into their disk-frame components.

While most works have focused on exploring $v_{\phi}$, as discussed in Section \ref{sec:methods:rotation_curves}, it is possible to extract information on $v_r$ and $v_z$ if a few assumptions are made. The largest of these is azimuthal symmetry, at least over the azimuthal region of the disk that is being investigated. Following the same procedure as described in Section~\ref{sec:methods:rotation_curves}, a more complex model including all three velocity components can be used,
\begin{equation}
v_0 = v_{\phi,\, proj} + v_{r,\, {\rm proj}} + v_{z,\, {\rm proj}} + v_{\rm LSR}
\end{equation}

By shifting the spectrum in each pixel by the projected disk rotation, accurate measurements of rotation curves can be performed (down to $\approx$ 10\,m\,s$^{-1}$, \emph{i.e.} a fraction of the spectral resolution of ALMA: \citealp{Teague2018, Teague2018b, Casassus2019b}). These methods revealed radial pressure gradients and meridional gas flows, likely driven by gaps carved in the gas surface density by Jupiter-mass planets in the disk of HD~163296 \citep{Teague2018, Teague2019}. Likewise, a  radial velocity component corresponding to  stellocentric accretion has been identified in the central cavity of HD\,135344B \citep[][]{Casassus2021}, amounting to $4\pm0.5$\% Keplerian, and the 3D velocity field in the gap separating the two rings follows similar radial profiles as the meridional flows seen in HD\,163296.

\subsection{Disambiguation with other physical mechanisms}

\begin{figure*}[!t]
  \centering
  \includegraphics[width=\hsize]{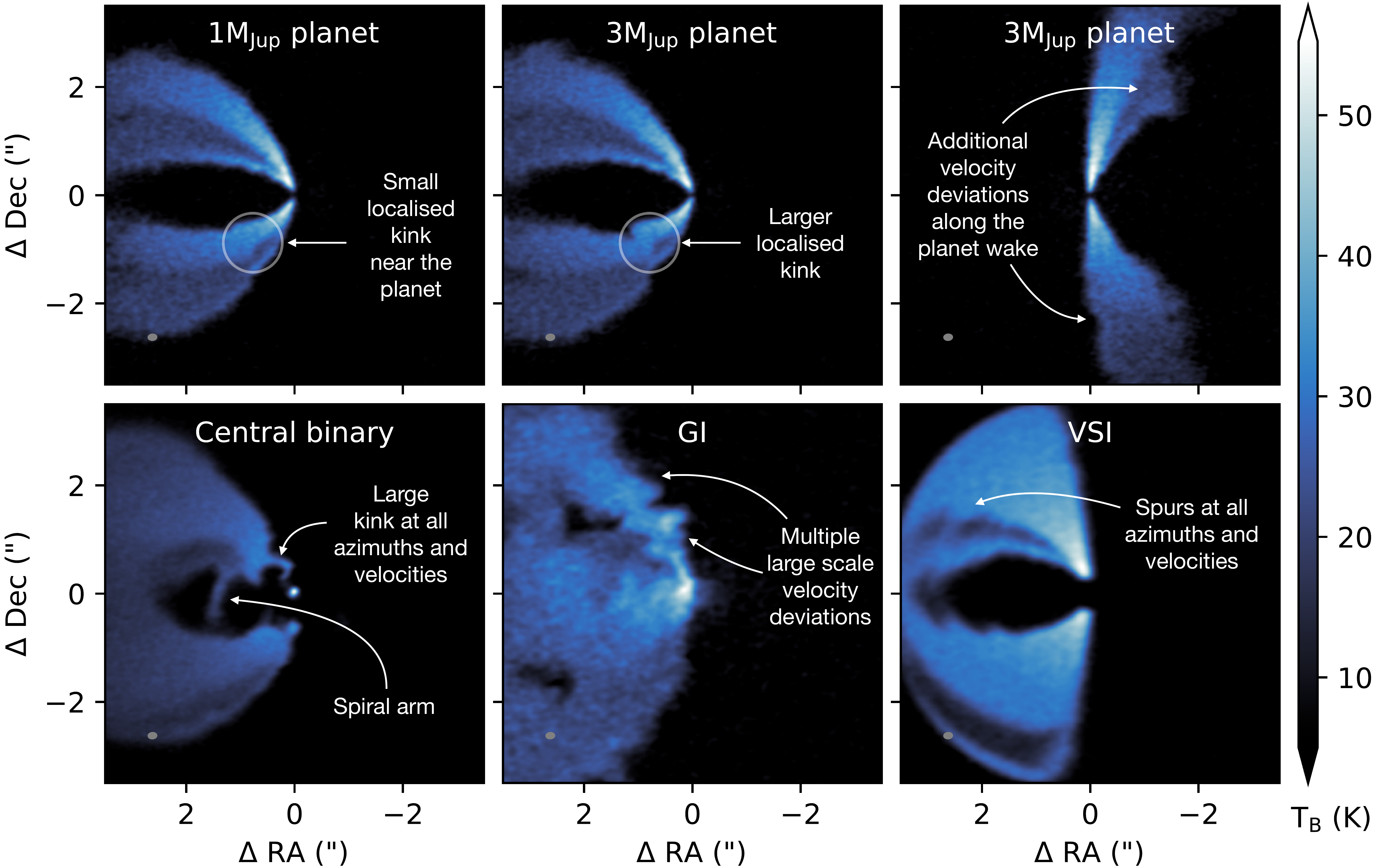}
  \caption{Synthetic $^{12}$CO channel maps for a sample of models. Planets produce signatures with amplitudes that depend directly on the planet mass (models adapted from \citealp{Pinte2018b}). The velocity deviations are stronger near the planet and produce a distinctive kink in the emission, but additional deviations can be detected along the planet wake if the planet is massive enough. Interactions with a central companion (Calcino et al., in prep.), GI \citep{Hall2020} or VSI \citep{Barraza-Alfaro2021} models predict distinct signatures that are observed over the whole disk surface. All the models have been processed with the same parameters in the ALMA simulator (10h on source, beam of 0.1", spectral resolution of 50\,m\,s$^{-1}$).}
\label{fig:models}
\end{figure*}

As discussed in the introduction, many mechanisms have been proposed to explain the observed dust structures seen in thermal continuum emission and scattered light. Most of these processes impact the gas as well as the dust, and will have corresponding signatures in kinematics. Kinematic predictions have only been made for a few types of models, but recent work highlights  whether and how we can expect to disentangle the signatures of planets from other physical processes in the next few years.

Figure~\ref{fig:models} illustrates the expected signatures in selected channel maps for a family of models, highlighting how the emission morphology differs: a central binary generates large kinks inside the cavity where the gas falls at near free-fall velocities and large scale spirals outside the cavity, gravitational instabilities produce the large scale ‘wiggles’ discussed in section~\ref{sec:GI}, while the vertical shear instability leads to ‘spurs’ of the order of 50\,m\,s$^{-1}$, that are visible at all azimuths in the disk if the inclination is low enough \citep{Barraza-Alfaro2021}. In contrast, planets produce strong deviations near the planet, or equivalently sign reversals in the rotation maps. Note that at high signal-to-noise, the planet wake can be detected across most of the disk \citep[e.g.][]{Bollati2021,Calcino2022}. The amplitude of the velocity deviations decreases with distance from the planet, but care must be taken when interpreting line observations as projection effects can impact the visibility of the kinematic signatures and part of the wake can be hidden.
These synthetic observations also demonstrate that for dedicated observations, with long integration times ($\gtrsim$ 10h), a visual inspection of the channel maps readily enables the detection of a 1M$_\mathrm{Jup}$ planet.

While the various models display different kinematic signatures, they also do share a lot a features in common. Pinpointing the physical origin of a given kinematic structure in the data, especially at the low level needed to try to detect a planet of one Jupiter mass or less, will require a deep understanding of all the processes at play.
For instance, while the planet wake is the most easily detectable near the planet, it extends over the whole disk, generating multiple low amplitude distortions in the channel maps \citep[][]{Bollati2021}. Theoretically, tracing the full spiral wake will help to validate the planet scenario \citep{Calcino2022}.  Note that the visibility of the kink or sign reversal associated with the planet wake depends strongly on azimuth. The full wake might only appear intermittently as a function of the azimuth in deep data sets. \cite{Rabago2021a} showed that the deviations from Keplerian velocity of the azimuthally averaged velocities are due to the gap profile itself, and not the planet. They suggest to combine both azimuthally averaged rotation maps, and localized velocity deviations to detect young planets in disks.

Most of these processes, and the resulting kinematic signatures, strongly depend on the underlying physical structure (temperature and density) of the disk. Additional constraints on these key parameters are critically required to test further the different models. Using multiple tracers sampling the vertical extent of the disk would be particularly useful to refine those constraints. This will also enable the mapping the gas flow in 3D, offering more diagnostic to disentangle between the various hypothesis. Most kinematic observations have focused on the brightest upper layer of the disk. With deeper observations, the signal-to-noise on the lower surface may become large enough to search for kinematic signatures on that surface, offering to a way test whether the structures are symmetric about the midplane.

\subsection{Caveats on the disk kinematics as a planet detection method}
\label{sec:caveats_planets}

While the current kinematic detections are well matched by models of embedded planets, there are several issues with the current studies that need to be addressed, or at least examined, in order to be more widely adopted as a new method to hunt for young planets.

The precise origin of the kink has not been studied in full detail yet, but it likely arises from the wake launched by the planet at the Lindblad resonances. This is confirmed by the semi-analytical work of \cite{Bollati2021} which can produce a kink by only including the planet wake, \emph{i.e.} without the presence of a gap (and hence pressure gradient), Hill sphere and horseshoe orbits. It remains unknown whether these extra components play a role in the observed kinematic signatures, in particular when observations with multiple lines probing a range of optical depths will become available.

As the flow of gas around a planet is mostly set by the planet mass, kinematic detections of embedded planets potentially allow for accurate estimates of the mass of planets still embedded in their parental disk. So far this has relied on matching a suite of hydrodynamical models with fixed parameter and a vertically isothermal structure \citep[e.g.][]{Pinte2018b,Pinte2019,Teague2018}. It not well understood how the estimated planet mass depends on the adopted disk models however. The work by \cite{Bollati2021} provides the first answers. First, because it potentially allows for a much faster exploration of the parameter space. And second, because analytical derivations reveal the meaningful parameters. As \cite{Bollati2021} showed that the kink is mostly created by the planet wake, it means that the relevant unit for the planet mass is the thermal mass $m_\mathrm{th} = (h_p/r_p)^3 M_*$.
To convert this constraint to an actual planet mass, an accurate estimate of $h_p$, and hence of the thermal disk structure, is critical. As discussed in sec.~\ref{sec:channel_maps}, observations of multiple molecular line provide a way to at least partially map the disk temperature structure. Progress in numerical methods and computing power now also offer the possibility to validate the adopted thermal structure by comparing the results from the vertically isothermal models to radiative-equilibrium hydrodynamics simulations \citep[e.g.][]{Pinte2019}, but this in turn depends on the adopted opacities.  Clearly more theoretical work is required to fully validate these methods. \cite{Cimerman2021} showed that the semi-analytical formulation by \cite{Bollati2021} may underestimate the velocity deviations in the planet wake and could benefit from a calibration based on numerical simulations in the linear region, near the planet. An exciting observational avenue to refine current planet mass estimate is to map a larger fraction of the spiral wake with deeper ALMA observations. As the velocity deviations near the planet scale on $M_p / (h_p/r_p)^3$, and the opening angle of the wake only depends on $h/r$, it should possible to disentangle between $M_p$ and $h/r$ if the wake is observed over a wider range of radii \citep[e.g.][]{Calcino2022}.

Semi-analytical formulations are also limited to planet masses below the thermal mass, \emph{i.e.} non-gap opening planets,  for which the linear and non-linear regimes of the wake propagation can be separated. The currently detected planets in HD~163296 and HD~97048 appears to slightly larger than the thermal masses, and more work is required to properly understand the propagation of the wakes in this regime.  Analytical developments have also only been developed in 2D so far, ignoring vertical temperature gradients, which can for instance lead to additional buoyancy spirals discussed in sec.~\ref{sec:spirals}. The generalisation of radiation hydrodynamics simulations, with realistic thermal structures, will be essential to establish more robust constraints on planet masses.

One final difficulty is to assess the statistical significance of the detections in individual channel maps. The detections to date have been performed by eye \citep{Pinte2018b,Pinte2019}. A more quantitative and robust procedure is needed, in particular to extend these detections to lower planet masses. Many parameters affect the kinematic signatures of an embedded planet: the disk thermal structure as discussed above, but also the orientation of the system and azimuth of the planet in the disk, as well as the observational setup (choice of lines, spectral resolution, integration time, $uv$ coverage and imaging algorithm). Additionally, observational biases, such as continuum subtraction and steep spatial gradients in line emission need also to be considered (sec.~\ref{sec:limitations}). \cite{Izquierdo2021} proposed a statistical framework (\texttt{discminer}) to detect and quantify the kinematic perturbations driven by a planet,  and to distinguish them from the signatures of a gap. Their algorithm successfully recovered a 0.3\,M$_\mathrm{Jup}$ planet at all azimuths from synthetic observations. This is a factor $\approx$ 10 lower in mass compared to the current detections. While this work only addresses part of the effects that may affect the kinematic signature of planet, it certainly offers an exciting strategy for detecting embedded planets in the next few years. Application of this framework on the disk surrounding HD~163296 recovered the velocity kink initially detected by \cite{Pinte2018b}, as well as a new planet candidate at 94\,au, \emph{i.e.} within a gap in dust continuum emission \citep{Izquierdo2022}.

\subsection{Criteria for a planet kinematic detection}

Given these limitations and caveats in both the data analysis tools, and predictions from numerical models, \cite{Pinte2020} proposed a set of criteria to be able to claim an embedded planet detection. These criteria were then extended and refined by \cite{Disk-Dynamics-Collaboration2020}. We update them here in light of recent results:
\begin{enumerate}
\item the detection of a \emph{localized} velocity disturbance (either in channel maps or in the rotation map), with potential additional kinematic signatures along the spiral wake,
\item the main velocity disturbance needs be to co-located with a gap, ideally detected in multiple tracers (sub-mm continuum, molecular emission or scattered light),
\item the detection of this velocity disturbance in multiple lines, ideally sampling the vertical structure of the disk.
\item the detection of the velocity disturbance needs to be resolved in velocity (e.g. detected in multiple independent channels)
\item a localized enhancement of the line broadening (ideally collocated with the velocity disturbance, but at least at the same orbital radius).
\end{enumerate}

The planet at 260\,au in the disk surrounding HD~163296 is currently the only detection that satisfy all these criteria \citep{Pinte2018b,Teague2018,Teague2019,Pinte2020,MAPS18,Izquierdo2022,Calcino2022}.
We note that HD~97048~b \citep{Pinte2019} has been added to the NASA's Exoplanet Database\footnote{\url{https://exoplanets.nasa.gov/exoplanet-catalog/7503/hd-97048-b/}}, although it currently only satisfy criteria \#1, 2 and 4, while the European exoplanet database\footnote{\url{https://exoplanet.eu/}} considers the two kinematic detections at 94\,au \citep{Izquierdo2022} and 260\,au in HD~163296 as confirmed, and lists HD~97048~b as a candidate.

\subsection{Constraints on planet formation}
\label{sec:planet_formation}

\begin{figure*}
  \includegraphics[width=\textwidth]{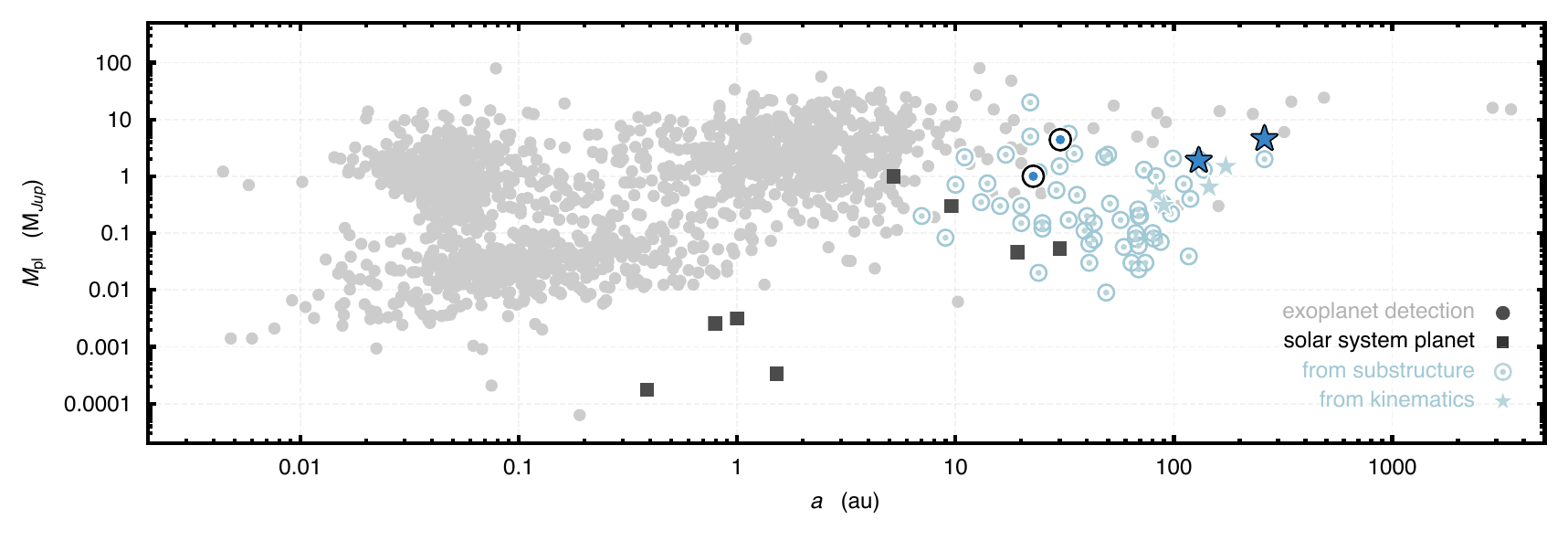}
  \caption{Current state of exoplanet demographics. Gray points show mature planets detected through typical means, while blue points highlight exoplanets, both confirmed and candidates, detected through sub-mm observations. Circles with dots represent planets detected through continuum substructures, while stars show planets inferred through kinematic means. Larger symbols denote the confirmed exoplanets: PDS~70b and PDS~70c, HD~97048b and HD~163296b.\label{fig:exoplanets}}
\end{figure*}

Kinematic observations allow for the detection of planets embedded in their natal disk, which had so far escaped detection with more classical techniques (Fig.~\ref{fig:exoplanets}). This offers an unique opportunity to observe planet formation in action. Knowing where planets are and how massive they are is critical to connect the disk physical and chemical properties to the planet formation process.

The origin of the currently detected planets, a few Jupiter mass planets at hundred of astronomical units from their central star, remains unclear (see chapter by Drazkowska et al.). The location of giant planets in the outer regions of disks is broadly consistent with gravitational instability \citep{Boss1997,Kratter2016}, but the typical planets formed via gravitational instabilities are expected to be $\approx$10M$_\mathrm{Jup}$ \citep{Forgan2011,Forgan2013,Forgan2018,Hall2017}.
On the other hand, the original core accretion framework \citep{Pollack1996} appears too slow to form planet embryos at hundred of astronomical units in less than a few million years, in particular because of the radial-drift barrier \citep[pebble migrates rapidly inwards due to gas drag, e.g.][]{Weidenschilling1977} and fragmentation barrier \citep[large particle shatter as their collision velocities increase with size, e.g.][]{Benz2000}. Major theoretical efforts have been dedicated to solve this acute dust growth timescale problem. The formation of a flat dust sub-disk, now observed with ALMA \citep[e.g.][]{Pinte2016,Louvet2018,Villenave2020}, provides the necessary conditions for the streaming instability to develop \citep{Youdin2005}. If the density in the dust layer become high enough, this may lead to clumping of pebbles \citep{Johansen2007a}, and subsequent formation of planetesimals \citep{Johansen2007b}. The growth of this planetesimal inside the pebble layer is then significantly faster, as the gas drag will increase the gravitational focusing of the pebbles, leading to ``pebble accretion'' \citep{Ormel2010,Lambrechts2012}. Thanks to these mechanisms, pebble accretion appears as a viable mechanisms to form planets at large separation on a million year timescale \citep{Johansen2017}. Kinematic detections of planets at shorter separations will provide tighter constraints on the respective roles of gravitational instabilities and pebble accretion.

The collocation of the kinematic signatures with the dust gaps (even if most detections to date are tentative) suggests that a significant fraction of the gaps are carved by planets. This would imply that planet formation is already largely advanced by the time we observed the disks. Planet and star formation may be simultaneous, as also suggested by the spectacular examples of [BHB2017]~1 \citep{Alves2020}, and the dust gaps detected in the disks of Class I protostars \citep{Sheehan2020,Segura-Cox2020}. If this is the case, planet formation might operate on an even shorter timescale ($\lesssim 1$\,Myr) than the 3-5\,Myr of the disk lifetime, and the initial conditions for planet synthesis models will need to be updated. If formed early, the planets detected to date may also have already undergone some outward migration, depending upon the initial profile of the disk, complicating even more the discussion of their origin. Comparison with signatures of migration in the dust \citep[e.g.][]{Perez2019b} may help identify planet migration.

Planet properties can also be discerned via their influence on the underlying dust and gas substructure (see chapters by Bae et al. and Lesur et al.). Most of the constraints are available from high resolution continuum observations. As such, large efforts have been invested in understanding the physics of dust gap opening \citep{Paardekooper2004,Fouchet2007,Fouchet2010,Dipierro2015,Dipierro2016,Rosotti2016,Dipierro2017,Dong2017b,Zhang2018a}, in particular with the distinction of two regimes depending on whether the planet is massive enough to open a gas gap \citep{Dipierro2016}. A key parameter is the Stokes number (ratio of the dust stopping time to the orbital time), which requires knowledge of the dust properties and background gas density profile. The assumed thermodynamics is also critical to estimate the gap profile \citep{Miranda2020}. All these quantities remain unfortunately poorly constrained, even though tremendous progress has been made in recent years \citep[see for instance the discussion by][ and chapter by Miotello et al.]{Andrews2020}. Kinematic observations can provide essential pieces of the puzzle, thus refining our models of planet-disk interactions and their impact on the dust distribution. As discussed in sec.~\ref{sec:caveats_planets}, the kinematics of the gas flows around the the planet mainly depend on the planet mass and thermal structure of the disk. If reasonable estimates for the temperature can be obtained, kinematic detections provide a direct handle on the planet mass. The associated dust structures in turn provides constraints on the Stokes number of the dust grains emitting the most at the considered wavelength. Using this method, \cite{Pinte2019} showed that the dust grains emission at 885\,$\mu$m in HD~97048 have a Stokes number of $\approx 10^{-2}$, suggesting that they might be porous aggregates (or that the gas density is severely underestimated). Additional studies on a sample of disks are critically needed to assess if this result is representative. This is critical to calibrate the dust gap/planet mass relations and push further comparisons between young planets as derived from dust gap with the population of mature exoplanets. The so far detected young planets occupy a distinct region from the bulk of the know exoplanets, in particular in terms of orbital parameters (Fig.~\ref{fig:exoplanets}). This mostly reflects the respective biases in the various planet detection methods. With current observations, it remains impossible to establish whether the populations of young and mature exoplanets are distinct \citep{Zhang2018a,Lodato2019}, and to assess the role of planet migration in a statistical way (see chapter by Paardekooper et al. for a more detailed discussion).

\section{Conclusions and future prospects}

ALMA and extreme adaptive optics system have transformed our view of disks and planet formation. Disks are highly structured, at a multitude of scales, and in three dimensions. The ubiquity of gaps, rings, azimuthal asymmetries and spirals in sub-millimeter continuum and near-infrared observations strongly suggests that planet-disk interactions play a significant role in shaping the disks. However, none of the alternative explanations, in particular (magneto\-)hydrodynamical instabilities, can be excluded. The origin of the sub-structures is a burning  open question. Are we seeing ``protoplanetary'' disks, in which the sub-structures are signposts of the planet formation process? or are we observing ``planet-hosting'' disk, where young, nearly formed planets are shaping the disks and creating sub-structure? The direct imaging of young giant planets in PDS~70
spectacularly demonstrated this hypothesis in one case. Could both mechanisms operate simultaneously, eventually leading to multiple generations of planets?

Although  molecular line observations cannot achieve the same spatial resolution as continuum observations, they offer the ability to map the gas flow associated with the dust sub-structures, opening a new unique window to probe the driving physical processes. The velocity dimension also offers a way to detected embedded planets, that would remain otherwise undetectable.
Disk kinematics is an emerging field. Results are still scarce, and are only showing the tip of the iceberg of discoveries ahead of us.
Several avenues will certainly need to be explored in the next decade.

Deep, high spectral and spatial resolutions line observations are critically needed. Not only they are required to detect the smallest deviations from Keplerian rotations, they are also essential to establish quantitative constraint on the disk density and temperature structures. Moving to longer integration times with ALMA will also allow us to push to smaller angular resolutions, allowing us to detect shorter period and smaller mass planets.

\cite{Perez2015} predicted that the main kinematic signature of an embedded planet would be the CPD, as it hosts the strongest deviation from the background Keplerian rotation.
Such signatures have not been detected to date. The detection of a CPD around PDS~70c in the sub-millimetre continuum \citep{Benisty2021} further motivates the search for the associated gas flows via deep high resolution line observations.
Longer observations will also open the prospect for  mapping the full extend of the planet wake, allowing us to break the degeneracy between planet mass and the thermal structure of the disk, which will lead to more accurate determinations of the planet masses.

Characterising the 3D flow of matter onto a planet will require combining multiple tracers that will probe the vertical extent of the disk. Such observations will provide critical information to understand the building of planet atmosphere, and on their composition. The C/O ratio is of particular importance as it can be measured in the atmospheres of exoplanets \citep{Madhusudha2019}. Recent progress has been made on mapping the C/O ratio in disk \citep{Bergin2016,Cleeves2018,Miotello2019,LeGal2019}. Finally, deep observations of multiple molecules might allow to trace the chemical signatures of planets \citep{Cleeves2015}.

Fundamental work on the theoretical side is required to aid the interpretation of observations. While there is a consensus on the global pattern of the flows generated by a planet in a disk, there is still much work to be done on understanding how they are affected by different assumptions, in particular on the disk thermodynamics. A better understanding of the kinematic signatures from a variety of non-planet related instabilities will help in understanding the respective roles of the mechanisms at play.
Better knowledge of planet signatures will allow us to detect smaller mass planets (more subtle deviations) with current data.

Statistically meaningful comparisons between the populations of young planets embedded in disks and mature exoplanets will require the search for planets in the inner 10\,au of the disks, which will remain beyond the reach of ALMA in molecular lines. The extremely large telescopes (ELTs), will allow us to probe the warm gas through ro-vibrational molecular transitions, as well as recombination lines such as Br$\gamma$ (for instance with the HARMONI integral field unit on the ESO/ELT). In conjunction, the extended missions of TESS and GAIA will grow more sensitive to longer period planets, starting to bridge the gap \citep[e.g.][]{Perryman2014}.

Finally, JWST and the  ELTs will offer unique opportunities to directly image the protoplanets responsible for the velocity deviations probed by ALMA.
A simultaneous detection of a velocity deviation (providing a mass estimate) and a direct image (giving a luminosity estimate) of the corresponding planet and/or associated circumplanetary disk  can help decide between the two major hypotheses for planet formation, as well as calibrate the planet formation models.
Whether planets are born in a ``cold'' (core accretion) or ``hot start'' (GI), or most likely with an intermediate ``warm start'' is a key constraint on planet formation models, and on the nature of the accretion onto forming planets. Embedded planets might remain undetectable by direct imaging in the near-infrared due to obscuration by the disk. The combination of ALMA and JWST, which opens high resolution imaging in the mid-infrared regime (where dust opacities are reduced compared to near-infrared wavelengths), will offer a unique opportunity to pave the way for comprehensive characterisation of the youngest exoplanets.

\bigskip

\textbf{Acknowledgements.}
We thank Felipe Alves, Himanshi Garg, Jane Huang, Teresa Paneque Carre\~no, and Gerrit van der Plas for sharing their reduced data sets.
ALMA is a partnership of ESO (representing
 its member states), NSF (USA) and NINS (Japan), together with NRC (Canada),
 MOST and ASIAA (Taiwan), and KASI (Republic of Korea), in cooperation with the
 Republic of Chile. The Joint ALMA Observatory is operated by ESO, AUI/NRAO and
 NAOJ. The National Radio Astronomy Observatory is a
facility of the National Science Foundation operated under cooperative
agreement by Associated Universities, Inc.
This work was performed on the ozSTAR national facility at Swinburne University
of Technology. ozSTAR is funded by Swinburne and the Australian Government's
Education Investment Fund. We thank the anonymous referee for insightful
comments and suggestions.
C.P. acknowledges funding from the Australian Research Council via
FT170100040 and DP180104235.

\bigskip
\textbf{Data.}
This work made use of the following ALMA data sets:\\
-- HD~163296: 2011.0.00010.SV
, 2013.1.00601.S,\\ DSHARP: 2016.1.00484.L, and MAPS: 2018.1.01055.L.\\
-- Elias 2-27: 2013.1.00498.S, 2016.1.00606.S and\\ 2017.1.00069.S,\\
-- [BHB2017]~1: 2013.1.00291.S,\\
-- RU Lup: 2018.1.01201.S and DSHARP: 2016.1.00484.L,\\
-- UX Tau: 2015.1.00888.S, and 2013.1.00498.S,\\
-- HD~142527: 2015.1.01353.S,\\
-- GG Tau: 2018.1.00532.S,\\
-- HD~97048: 2016.1.00826.S.

\bigskip
\textbf{Software.}
This work made use of the following software:\\
-- \texttt{AstroPy} \citep{astropy:2013,astropy:2018},\\
-- \texttt{CASA} \citep{McMullin07},\\
-- \texttt{matplotlib} \citep{Hunter:2007},\\
-- \texttt{eddy} \citep{eddy},\\ \url{https://github.com/richteague/eddy},\\
-- \texttt{casa\_cube}\\ \url{https://github.com/cpinte/casa_cube},\\
-- \texttt{mcfost} \citep{Pinte2006,Pinte2009},\\
 \url{https://github.com/cpinte/mcfost},\\
-- \texttt{pymcfost},\\ \url{https://github.com/cpinte/pymcfost},\\
-- \texttt{phantom} \citep{Price2018},\\ \url{https://github.com/djprice/phantom},\\
-- \texttt{splash} \citep{splash},\\ \url{https://github.com/danieljprice/splash},\\
-- \texttt{FARGO-3D} \citep{Benitez-Llambay2016},\\ \url{http://fargo.in2p3.fr/}.\\

We discussed a few additional codes that were specifically developed to perform kinematic studies:\\
-- \texttt{DiskJockey} \citep{Czekala2015}, \\ \url{https://github.com/iancze/DiskJockey}, \\
-- \texttt{pdspy}, \url{https://github.com/psheehan/pdspy},\\
-- \texttt{ConeRot},\\ \url{https://github.com/simoncasassus/ConeRot},\\
-- \texttt{bettermoments} \citep{bettermoments},\\ \url{https://github.com/richteague/bettermoments},\\
-- \texttt{GoFish} \citep{gofish}, \\ \url{https://github.com/richteague/gofish},\\
-- \texttt{discminer} \citep{Izquierdo2021},\\
-- \texttt{CO\_layers} \citep{Pinte2018a}\\ \url{https://github.com/cpinte/co_layers},\\
-- \texttt{disksurf} \citep{disksurf}\\ \url{https://github.com/richteague/disksurf}.\\

\bigskip

\bigskip
\parskip=0pt
\bibliographystyle{pp7.bst}
\bibliography{biblio1.bib,biblio2.bib}
\end{document}